\documentclass[usenatbib,usegraphicx]{mn2e}

\usepackage{amssymb}
\usepackage{ulem}
\usepackage{textcomp}

\begin{document}

\title[Lick-index diagnostic of stellar populations]{Lick-index entanglement 
and biased diagnostic of stellar populations in galaxies\thanks{Also based on observations 
made at the Observatorio Astron\'omico ``G. Haro'' of INAOE, Cananea (Mexico)}}

\author[Buzzoni]{Alberto Buzzoni\\
INAF - Osservatorio Astronomico di Bologna, Via Ranzani 1, 40127 Bologna,
Italy\\
{\sf alberto.buzzoni@oabo.inaf.it}
}


\date{Received ; Accepted}

\maketitle

\label{firstpage}

\begin{abstract}

The Lick-index spectrophotometric system is investigated in its inherent statistical and 
operational properties to ease a more appropriate use for astrophysical studies.
Non-Gaussian effects in the index standardization procedure suggest
that a minimum S/N ratio has to be reached by spectral data, such as $S/N \ga 5$~px$^{-1}$ 
for a spectral resolution $R \sim 2000$. In addition, index (re-)definition in terms of narrow-band 
``color'' should be preferred over the classical pseudo-equivalent width scheme. 

The overlapping wavelength range among different indices is also an issue, 
as it may lead the latter ones to correlate, beyond any strictly physical relationship.
The nested configuration of the Fe5335, Fe5270 indices, and the so-called ``Mg complex''
(including Mg$_1$, Mg$_2$ and Mgb) is analysed, in this context, by assessing the 
implied bias when joining entangled features into ``global'' diagnostic meta-indices, like the 
perused [MgFe] metallicity tracer.
The perturbing effect of [OIII]$_{5007}$ and [NI]$_{5199}$ forbidden gas emission on Fe5015 and 
Mgb absorption features is considered, and an updated correction scheme is proposed when
using [OIII]$_{5007}$ as a proxy to appraise H$\beta$ residual emission. When applied to present-day
elliptical galaxy population, the revised H$\beta$ scale leads, on average, to 20-30\% younger 
age estimates.

Finally, the misleading role of the christening element in Lick-based chemical analyses 
is illustrated for the striking case of Fe4531. In fact, while Iron is nominally the main 
contributor to the observed feature in high-resolution spectra, we have shown that the Fe4531 
index actually maximizes its responsiveness to Titanium abundance.
\end{abstract}


\section{Introduction}

Since its very beginning, the Lick-index system was intentionally conceived as a tool 
to characterize elliptical galaxies and other old stellar systems 
\citep{faber77,faber85,davies87,gorgas90}. 
Only a few relevant absorption lines were originally taken into account, 
at the sides of the strong Magnesium feature around 5175~\AA, 
which blends the molecular hydride MgH and the atomic doublet Mgb \citep{mould78}.
Together with the 4000~\AA\ break, the Mg ``valley'' has since long been recognized 
as the most outstanding pattern in the optical spectrum of galaxies 
\citep{ohman34,wood63,spinrad65}.

The immediate interest for galaxy investigation directly drove a parallel effort
to extend the index database also to cool stars \citep[e.g.][]{burstein84} 
for their evident link with the study of old galaxies through population  
synthesis applications. At the same time, this also led to a more definite
settlement of the whole index system, greatly extended by new additions from the Lick 
Image Dissector Scanner (IDS) data archive including all the relevant absorption features 
in the optical wavelength range comprised between the 4000~\AA\ break and the Balmer 
$H\alpha$ line \citep{gorgas93,worthey94b}.
The original emphasis to old stellar populations extended then to hot and warm 
stars \citep{worthey97} to account for ongoing star formation scenarios, as in 
late-type galaxies \citep[e.g.][]{molla99,walcher06,perez09}.

In more recent years, other spectroscopic methods have been complementing the original 
Lick approach, aimed at tackling galaxy properties at a finer or coarser detail 
level. High-resolution spectroscopy (say at the typical resolving power, $R \ga 2000$ 
or so, of current extragalactic surveys) certainly added valuable clues, on this 
line. On the opposite side, a panchromatic analysis relying on low-resolution fitting 
of galaxy SED (often assembled by straight flux conversion of integrated broad-band
magnitudes) also became an increasingly popular way to deal with massive 
wealth of galaxy data, in order to characterize redshift distribution and other 
outstanding evolutionary features on a statistical basis 
\citep[see][for excellent reviews on these subjects]{walcher11,conroy13}.

For several reasons, however, none of these techniques is free from some limitations. 
From its side, high-resolution spectroscopy may be constrained by the intervening role 
of galaxy dynamics; at a typical velocity dispersion of $\sigma_v \simeq 300$~km~s$^{-1}$,
the induced line broadening limits the effective resolving power at $R \la 1000$.
On the other hand, global fitting of galaxy SED carries on all the inherent uncertainty 
in absolute flux calibration of the observations, especially when matching data from
different observing sources and over distant wavelength ranges. In addition, to quote 
\citet{walcher11}, we have to remind that, as ``the uncertainties are dominated by 
the uncertainties in the SED modeling itself, thus one has to be very cautious about 
the interpretations when selecting samples where a specific type of model is preferred.''

\begin{figure*}
\centerline{
\includegraphics[width=0.6\hsize,clip=]{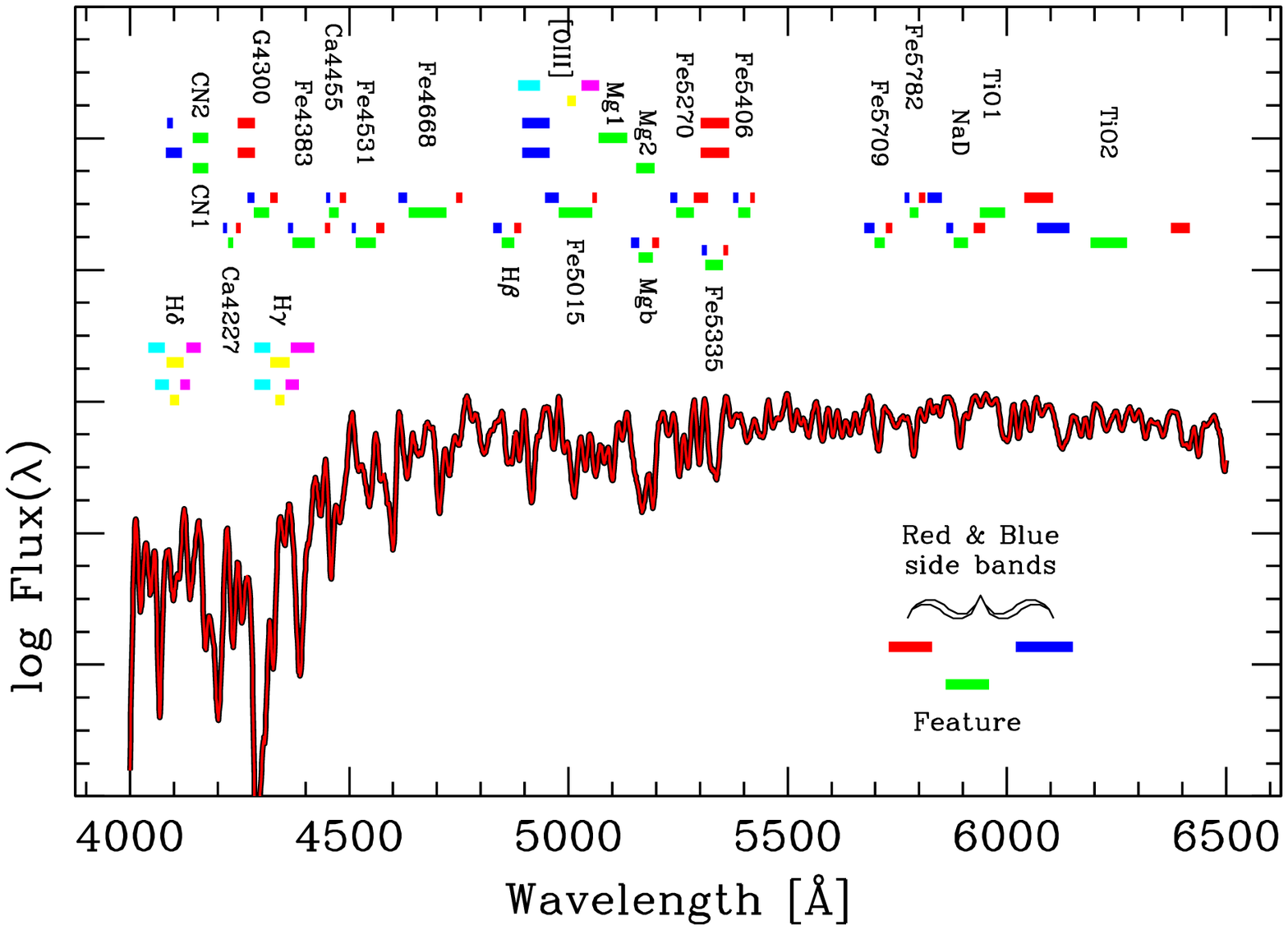}
}
\caption{
A global view of the Lick-index wavelength distribution. The original set of 21 indices \citep{worthey94b}
has been complemented by the four \citet{worthey97} new Balmer indices (namely, H$\delta$ and H$\gamma$, in their
``wide'' and ``narrow'' version) and the \citet{gonzalez93} [OIII] index, as labelled in the plot. 
The index map is superposed to a synthetic stellar spectrum with $[T_{\rm eff},\log g, Z] = [4750~{\rm K}, 2.00, Z_\odot]$
from the {\sc Bluered} theoretical library \citep{bertone04,bertone08}, degraded to match the Lick spectral
resolution of 8.5~\AA\ FWHM.
}
\label{f1}
\end{figure*}

For all these reasons, narrow-band spectrophotometry still remains, in most cases, the elective 
(and often the only viable) tool to assess in finer detail the evolutionary status of galaxies 
and other unresolved stellar aggregates both in local and high-redshift environments 
\citep[see][for recent contributions]{bernardi06,carson10,dobos12}.
Such a broad range of applications urges, however, a more careful 
in-depth analysis of the inherent properties of the Lick system in order to
enable its appropriate use in the investigation of single stars and integrated stellar 
systems. This motivated the present work.

When dealing with Lick indices, two intervening difficulties may in principle affect 
our conclusions. A first, and often underrated issue (discussed here in Sec.~2) deals with 
the index {\it standardization},
that is with the process to make index measurements, from different sources or observing
circumstances, to consistently compare eachother. Quite importantly, this process
should not be confused with the {\it calibration} procedure, as the first requires 
{\it ab initio} that index definition could lead anytime to a ``fair'' statistical 
realization of a normal statistical variable according to the Central Limit Theorem
\citep[see, e.g.][for a more rigorous theoretical settlement of the problem.]{sachs84}
While a satisfactory attention is devoted in the relevant literature to suitably calibrate
observing data \citep{worthey94b,trager98, cardiel98}, this cannot assure, by itself, that a 
``standard'' settlement of the derived index estimates is always achieved.

The latter feature we have to remind is that indices may not be completely 
{\it independent} each one and, how we will show in Sec.~3, some degree of redundancy occurs 
in many cases.
For instance, sometimes two indices may share the same pseudo-continuum window,
or two feature windows are partly or fully superposed etc.. Even more sneaky, some 
redundancy may also appear when, under certain conditions, the same chemical element
that constrains the main feature also affect the surrounding pseudo-continuum,
thus masking or vanishing the index sensitivity to its abundance.
When changing temperature or other fundamental parameters of stars, a ``hidden'' 
chemical element may abruptly appear at some point with its lines, and superpose 
to the other absorption features affecting the corresponding indices. This is,
for instance, the classical case of the onset of TiO molecular bands in cool stars 
perturbing, among others, the Mg$_2$ and Fe5270 indices \citep{buzzoni92,tantalo04,buzzoni09a}.

A further range of problems deal with more physical environment conditions
of stellar populations. Most importantly, a natural effect of fresh star formation on the 
integrated spectrum of galaxies is line emission driven by the residual interstellar gas. 
Along the Lick-index range, this results in enhanced Hydrogen Balmer lines and, depending 
on the overall thermodynamical conditions, also in [OIII] forbidden emission 
at 4363, 4959 and 5007~\AA\ \citep[see][for excellent explanatory examples]{kennicutt92}. 
On the other hand, Oxygen and $H\beta$ emission is triggered even at the vanishing star 
formation rates observed in most elliptical galaxies 
\citep{caldwell84,phillips86,bettoni87,volkov90,sarzi10,crocker11} due, at the very least, to
the contribution of the planetary nebulae \citep[see][for an example]{kehrig12}, and this 
could easily bias our conclusions if we use $H\beta$ (absorption) strength to constrain 
galaxy age.

In this paper we will carry on our analysis from two independent points of view, starting
from the expected behaviour according to index properties and checking theoretical
predictions by means of a tuned set of spectroscopic observations for a sample of twenty 
elliptical galaxies.
Our relevant conclusions, and a reasoned summary of the the manifold variables at work 
is proposed in Sec.~4 of the paper.

\section{Index technicalities}

Up to the 70's, low-resolution sampling of galaxy and stellar energy distribution 
was mainly pursued by narrow-band photometry. Popular multicolor systems included,
for instance, those of \citet{stromgren56}, the DDO system of \citet{mcclure68}, 
the multicolor set of \citet{spinrad69} or the 10-color system of \citet{faber73}.
In this framework, the Lick system stands out for its innovative approach to the problem,
tackled now from a purely spectroscopic point of view.

Indeed, such a new ``spectroscopic mood'' is inherent in the Lick-index definition. 
The procedure applies in fact to low-resolution ($R \sim 500$) spectra and aims at 
deriving the strength of a number of absorption features through a measure of their
pseudo-equivalent width. Indices are therefore measured in \AA, maintaining
the magnitude scale only for those few cases of molecular bands \citep{worthey94b}. 
The original set of 21 indices was defined by \citet{worthey94b}
and slightly adapted to more suitably fit with extragalactic targets by \citet{trager98}.
A complement to include the full Balmer line series in the optical wavelength range
led \citet{worthey97} to add four new indices, that match $H\gamma$ and $H\delta$ lines
with ``narrow'' and ``wide'' windows. With these additions, the full Lick system consists
of 25 indices.
Apart, it is worth including in our analysis also an extemporary addition to the standard
set due to \citet{gonzalez93}, in order to account for [OIII] emission in the
spectrum of star-forming galaxies. Note that this is the only index aimed at
measuring an emission feature (see Fig.~\ref{f1}).

According to \citet*{worthey94b} original precepts, the key quantity of any Lick index 
is the ratio
\begin{equation}
{\cal R} = \overline{\big({f\over f_c}\big)} = {1\over w}\ \int_w \left({f \over f_c}\right)_\lambda d\lambda.
\label{eq:r}
\end{equation}
$\cal R$ is therefore an average, along the $w$-wide feature window, 
of the apparent flux density, $f(\lambda)$, normalized to the (pseudo)-continuum level, 
$f_c(\lambda)$, linearly interpolated at the same wavelength. Two adjacent 
bands, on both side of the feature, provide the baseline for the interpolated process.
With these prescriptions, the index then directly follows either as
\begin{equation}
I({\rm \AA}) = w\ \big(1-{\cal R}\big)
\label{eq:ia}
\end{equation}
or
\begin{equation}
I({\rm mag}) = -2.5\ \log {\cal R}
\label{eq:imag}
\end{equation}
depending whether we want to express it in \AA, as pseudo-equivalent width,
or in magnitude scale.
Clearly, both definitions are fully equivalent and easily reversible as
\begin{equation}
I({\rm \AA}) = w\ \big[1-10^{-0.4\ {\rm I(mag)}}\big]
\label{eq:4}
\end{equation}
or
\begin{equation}
I({\rm mag}) = -2.5\ \log \big[1-{{\rm I(\AA)}\over w}\big],
\end{equation}
and the use of \AA\ or magnitudes is just a matter of ``flavor'', as we have been
mentioning above (see, for instance, \citealt{brodie90}, \citealt{colless99}, or 
\citealt{sanchez06a} for alternative non-standard notations). As a rule, both 
eq.~(\ref{eq:ia}) and (\ref{eq:imag}) are positive for absorption features and
negative for emission ones.

\subsection{Standardization constraints}

In spite of its established popularity and extensive use in a broad range of stellar and
extra-galactic studies, a well recognized drawback of the Lick system is its ambigous 
standardization procedure. A meticulous analysis in this sense has been attempted by
\citet{trager98} to account for the  macroscopic limits of the IDS database due to the
somewhat fickle technical performance of the instrument. Some inconstancy in the spectrum 
wavelength scale is, for instance, a recognized issue, and even the spectral resolution 
\citep[about 8.5~\AA\ FWHM, see][]{worthey97} cannot be firmly established due to slightly 
unpredictable behaviour of the instrument output with time \citep{worthey94b}. In addition, 
and quite importantly, one has to notice that the whole IDS data set is in fact {\it not} 
flux-calibrated \citep{worthey94b}, thus taking over the instrument (slightly variable) 
response curve.
For all this, any comparison between external data sources in the literature must forcedly 
rely, {\it as a minimum}, on the observation of a common grid of standard stars 
\citep[e.g.][among others]{worthey94b,gorgas93,buzzoni92,buzzoni94,buzzoni01}.

A few attempts have been made, in the recent literature, to refine the index estimates from
the operational point of view to secure on more solid bases the observed output.
The work of \citet{rogers10} is certainly the most relevant one, in this sense, as it 
especially focus on the problem of a fair settlment of the pseudo-continuum level 
(through the so-called ``boosted median continuum'' method) to derive 
a more confident feature strength, also overcoming those recognized cases of unphysical 
output (for example with nominally ``emission'' indices tracing, in fact, absoption features etc.).

As a matter of fact, any more or less elaborated empirical matching procedure, however, 
cannot overcome a more subtle but inherent difficulty in the standardization process of 
the Lick system.
In fact, the way one leads to compute $\cal R$ according to eq.~(\ref{eq:r}) is a direct
``heritage'' of high-resolution spectroscopy, where the measure is always carried out
on preliminarily continuum-rectified spectra. However, this is clearly {\it not} the case 
for the (unfluxed) Lick spectra, where possibly steep continuum gradients are in principle an issue.
According to statistics theory, the consequence is that {\it ratio $\cal R$ may display 
strongly non-Gaussian properties}
\citep[see][and especially Fig.~1 therein for an eloquent example]{marsiglia64}
making the analysis of its distribution a hard (and in some cases hopeless) 
task. In other words, this implies that, in general, any effort to reproduce the system 
may not obey the Central Limit Theorem so that its convergence is not assured 
{\it a priori} \citep{marsiglia65}.

In this framework, it is therefore of central interest to explore the real contraints
that make $\cal R$ behave normally. For this, let us briefly consider the more general 
problem of the stochastic behaviour of a statistical variable $r$ being the ratio of two 
normal variables $x$ and $y$, namely $r = y/x$. It has been firmly demonstrated 
\citep{geary30,hinkley69,hayya75} that $r$ tends to behave like $y$ if the range of 
variation of $x$ (that is $dx/\overline{x}$), approaches zero and the $x$ variable is 
``unlikely to assume negative values'' \citep{geary30}. 
If this is the case, then $x$ tends to behave as a (positive) ``constant'' and 
$xr$ would resemble the $y$ distribution \citep{hayya75}. Under these assumptions,
the $r' = \overline{y}/\overline{x}$ ratio will display a normal distribution
and its average $\overline{r'}$ tends to be an unbiased proxy of $\overline{r}$.
In our terms, therefore, the crucial constraints that ensure both index Gaussianity and 
its fair reproducibility directly deals with two issues: from one hand we need a
``small'' range of variation for $df_c/\overline{f_c}$; on the other hand, we need
to assess the condition for which $\overline{r'} \to \overline{r}$.

To further proceed with our analysis, let us preliminarily assume the relative variation of 
$f_c$ {\it to be} ``small'', indeed, so that we are allowed to set up a linear expansion for $f_c$ 
around $\lambda_o$, the central wavelength within the feature window of width $w$:
\begin{equation}
{1\over f_c} = \big[\ {1\over \overline{f_c}} -{1\over \overline{f_c}^2}\ {\partial f_c\over \partial \lambda}\ (\lambda-\lambda_o) \big]
= {1\over \overline{f_c}}\ \big[1- {{\alpha (\lambda-\lambda_o)}\over \overline{f_c}}\big].
\label{eq:centroid}
\end{equation}
In previous equation, $\alpha = \partial f_c /\partial \lambda$ and, by definition, 
$({\alpha w}) = df_c$ is the maximum excursion of $f_c(\lambda)$ within the feature window.

By implementing eq.~(\ref{eq:centroid}) into eq.~(\ref{eq:r}) we have
\begin{equation}
{\cal R} = {1\over \overline{f_c}} \big[ {{\int_w f(\lambda) d\lambda}\over w}
- {\alpha\over {w \overline{f_c}}}\int_w f(\lambda)(\lambda-\lambda_o) d\lambda \big].
\end{equation}
If we now multiply and divide by $w \int f(\lambda) d\lambda$ the last term of the equation,
with little arithmetic we can re-arrange the equation and write
\begin{equation}
{\cal R} = {\cal R'} \big[ 1 - {{\alpha w} \over \overline{f_c}}
{{\langle\delta \lambda\rangle} \over w}\big] = {\cal R'} \big[ 1 - 
{{\langle\delta \lambda\rangle} \over w} \left({{df_c} \over \overline{f_c}}\right)^{\rm max}\big] 
\label{eq:rr1}
\end{equation}
being
\begin{equation}
{\cal R'} = \big({\overline{f} \over \overline{f_c}}\big), 
\label{eq:r1}
\end{equation}
and $\langle\delta \lambda\rangle/w$ the relative displacement, within the feature
window, of the line centroid with respect to $\lambda_o$, in consequence of symmetry departure due
to the spectral slope.
According to eq.~(\ref{eq:rr1}), therefore, ${\cal R'}$ tends to become a fair proxy of ${\cal R}$ if 
\begin{equation}
\Big| 1-{{\cal R}\over {\cal R'}}\Big| \la \left( {{\langle \delta \lambda\rangle} \over w}\right) \left( {{df_c} \over \overline{f_c}}\right)^{\rm max}.  
\label{eq:rg}
\end{equation}

The maximim relative excursion $(df_c /\overline{f_c})^{\rm max}$ is, evidently, the key figure in
this process as it directly constrains both r.h. factors in eq.~({\ref{eq:rg}).
It may be wise to assess the allowed range of this parameter in terms of the stochastic fluctuation range 
of $f_c$ within the same window, the latter being directly related to the signal-to-noise ratio 
($S/N$) of our data, so that we can set 
\begin{equation}
\big({df_c \over \overline{f_c}}\big)^{\rm max}  \la\ k \big[{\sigma(f_c) \over \overline{f_c}}\big] = 
{k \over S/N},
\label{eq:fmax}
\end{equation}
with $k$ a tuning factor.
We will show, in the next section, that a minimum signal-to-noise ratio of $S/N \sim 20$ is required 
for ${\cal R'}$ to behave normally \citep{hayya75}. With this figure, if we  choose to tentatively
accept a pseudocontinuum excursion within a $\pm 3\sigma$ range of the random noise, then
$k \la 6$ and $(df_c /\overline{f_c})^{\rm max} \la 0.3$, so that one could actually verify 
numerically that 
\begin{equation}
\Big| 1-{{\cal R}\over {\cal R'}}\Big| \la 0.01.  
\label{eq:rg2}
\end{equation}
With the previous assumptions, this means that, within a 1\% accuracy, the ratio ${\cal R'}$  
eventually becomes an unbiased estimator of ${\cal R}$, thus assuring ${\cal R}$ to behave 
normally, too.
Our first conclusion is therefore that {\it \citet*{geary30} prescriptions require pseudo-continumm
not to exceed a $\pm 30$\% relative variation within the feature window for any Lick index to
obey the Central Limit Theorem, and display a Gaussian distribution.}

\subsection{Random-error estimate}

A range of more or less entangled attempts have been proposed across the literature to
derive a general and self-consistent estimate of the statistical uncertainty
to be related to narrow-band indices 
\citep{rich88,brodie90,carollo93,gonzalez93,cardiel98,trager98}.
Yet, this effort does not seem to have led to any firm and 
straightforward pathway to assess the index random-error estimate.
Quite different formal approaches can be found among authors depending 
whether the intervening noise sources (including photon statistics,
flat-fielding procedure, sky subtraction etc.) are considered individually or
to a somewhat aggregated level in the analysis. In addition, as we have been discussing
in previous section, with the Lick system one has to deal with the duality of index definition, 
sometimes seen in terms of pseudo-colors and sometimes in terms of pseudo-equivalent 
widths.

In its general form, by simple differentiation of eq.~(\ref{eq:ia}) and (\ref{eq:imag}), 
the expected internal uncertainty of a Lick index expressed in magnitude is
\begin{equation}
\sigma(I)_{\rm mag} = 1.08 \big[{\sigma({\cal R})\over {\cal R}}\big].
\label{eq:ermag}
\end{equation}
On the contrary, if we chose to express the index in pseudo-equivalent
width, then
\begin{equation}
\sigma(I)_{\rm{\AA}} = \big[{\sigma({\cal R})\over {\cal R}}\big](w-I_{\rm{\AA}})
\label{eq:erang}
\end{equation}

In both cases, the crucial quantity that we have to assess is the $\sigma({\cal R})/{\cal R}$
ratio.
A useful contribution to settle the problem has come from
\citet{vollmann06}, dealing with the error estimate in equivalent-width measurements 
from high-resolution spectroscopy.
Under many aspects, their results directly apply to our framework providing the caveats
of previous section to be properly accounted for in our analysis and to observe at 
``sufficient S/N within the continuum'' \citep{vollmann06}. 
If the ${\cal R}'$ ratio could be an effective proxy of ${\cal R}$, then
the variance of ${\cal R}$ directly derives from error propagation as
\begin{equation}
\sigma^2({\cal R}) = \big[{\partial {\cal R} \over \partial f}\sigma(f)\big]^2+\big[{\partial {\cal R} \over \partial f_c}\sigma(f_c)\big]^2,
\end{equation}
which implies
\begin{equation}
\sigma^2({\cal R}) = \big({1\over f^2_c} \sigma^2_f\big) + \big({f^2 \over f^4_c} \sigma^2_{fc}\big) =
{f^2\over f^2_c}\ {\sigma^2_f\over f^2} + {f^2 \over f^2_c} \ {\sigma^2_{fc} \over f^2_c}.
\end{equation}
In case a similar S/N ratio along the relevant range of our spectrum can be assumed, 
so that $({\sigma(f)/f}) \sim ({\sigma(fc)/f_c})$, a final compact form can be achieved
for previous equations such as 
\begin{equation}
{\sigma({\cal R}) \over {\cal R}} \simeq {\sigma({\cal R}) \over {\cal R}'}= {\sqrt{2}\over (S/N)_{idx}}.
\label{eq:gr}
\end{equation}

Quite importantly, note that the value of $S/N$ in eq.~(\ref{eq:gr}) refers to the
{\it integrated} figure along the feature and continuum windows. Providing to observe 
a spectrum with an original $(S/N)_{obs}$ ratio {\it per pixel} and a dispersion 
$\theta$ \AA~px$^{-1}$, the resulting $(S/N)_{idx}$ by sampling flux along 
a window $w$ \AA\ wide is, of course,
\begin{equation}
(S/N)_{idx} = (S/N)_{obs} \sqrt{{w \over \theta}}
\label{eq:17}
\end{equation}

\begin{table}
\caption{Harmonic average$^{(a)}$ (${\cal W}$), feature (w) and continuum (W) window width for the extended set of Lick indices}
\label{t1}
\begin{tabular}{lrrrlrrr}
\hline\hline
Index & ${\cal W}$ &  w & W & Index & ${\cal W}$ & w & W \\
 & \multicolumn{3}{c}{\hrulefill~~[\AA]~~\hrulefill}  & & \multicolumn{3}{c}{\hrulefill~~[\AA]~~\hrulefill}  \\
\hline
 H$\delta_W$ & 50.1 & 38.8 &  70.6 & $[{\rm OIII}]$ & 32.7 &  20.0 &  90.0  \\ 
 H$\delta_N$ & 30.5 & 21.3 &  53.8 &  Fe$_{5015}$ & 54.6 &  76.3 &  42.5  \\ 
      CN$_1$ & 48.2 & 35.0 &  77.5 &	   Mg$_1$ & 86.1 &  65.0 & 127.5  \\ 
      CN$_2$ & 42.0 & 35.0 &  52.5 &	   Mg$_2$ & 63.8 &  42.5 & 127.5  \\ 
 Ca$_{4227}$ & 15.0 & 12.5 &  18.8 &	   Mg$_b$ & 33.1 &  32.5 &  33.7  \\ 
  G$_{4300}$ & 33.7 & 35.0 &  32.5 &  Fe$_{5270}$ & 43.4 &  40.0 &  47.5  \\ 
 H$\gamma_W$ & 58.6 & 43.8 &  88.8 &  Fe$_{5335}$ & 27.8 &  40.0 &  21.2  \\ 
 H$\gamma_N$ & 31.9 & 21.0 &  66.3 &  Fe$_{5406}$ & 24.0 &  27.5 &  21.3  \\ 
 Fe$_{4383}$ & 32.5 & 51.3 &  23.8 &  Fe$_{5709}$ & 29.1 &  23.8 &  37.5  \\ 
 Ca$_{4455}$ & 23.1 & 22.5 &  23.7 &  Fe$_{5782}$ & 21.7 &  20.0 &  23.7  \\ 
 Fe$_{4531}$ & 35.1 & 45.0 &  28.8 &	   NaD    & 36.3 &  32.5 &  41.0  \\ 
 Fe$_{4668}$ & 47.3 & 86.3 &  32.5 &  TiO$_1$	  & 72.3 &  57.5 &  97.5  \\ 
    H$\beta$ & 31.6 & 28.8 &  35.0 &  TiO$_2$	  & 97.0 &  82.5 & 117.5  \\ 
\hline\hline
\noalign{$^{(a)}$ According to eq.~(\ref{eq:d}).}
\end{tabular}
\end{table}

In their work, \citep{hayya75} set the limit of general validity of this formal approach through 
a detailed Monte Carlo analysis. Converting their results to our specific framework (see, 
in particular, their Fig.~1), we conclude that eq.~(\ref{eq:gr}) is a robust estimator 
of the relative standard deviation of ${\cal R}$ providing to work with $(S/N)_{idx} > 20$.

On the basis of our discussion we can eventually re-arrange previous eq.~(\ref{eq:ermag}) 
and (\ref{eq:erang}) in their final form, respectively:
\begin{equation}
\sigma(I)_{\rm mag} = {\sqrt{2\theta\over {\cal W}}}\ \ {1.09~~~~\over (S/N)_{obs}}, 
\label{eq:smag}
\end{equation}
and
\begin{equation}
\sigma(I)_{\rm \AA} = {\sqrt{2\theta \over {\cal W}}}\ \ {w\over (S/N)_{obs}}\ \big(1-{I_{\rm\AA}\over w}\big).
\label{eq:sang}
\end{equation}
For ``shallow'' lines (that is those with $[I({\rm \AA})/ w] \ll 1$), we can simply neglect the last 
factor in previous equation,  thus leading to an even more compact (though slightly overestimated) 
form for the standard deviation.

In both eq.~(\ref{eq:smag}) and (\ref{eq:sang}), ${\cal W}$ is the harmonic average of the feature 
($w$) and continuum ($W$) window widths:
\begin{equation}
{2\over {\cal W}} = \big({1\over w} + {1\over W}\big).
\label{eq:d}
\end{equation}
For reader's better convenience, the value of $\cal W$, together with the feature and the full (i.e.\ red+blue)
continuum windows for the extended set of Lick indices is summarized
in Table~\ref{t1}.

Taking eq.~(\ref{eq:smag}) and (\ref{eq:sang}) as a reference,
in Fig.~\ref{f2} we compare the observed standard deviation of primary Lick stellar 
calibrators, according to \citet[][see, in particular, Table 1 therein]{worthey94b} 
and \citet{worthey97} with our expected error distribution for a spectrum of 
$(S/N)_{\rm obs}=30$.

\begin{figure}
\centerline{
\includegraphics[width=0.7\hsize,clip=]{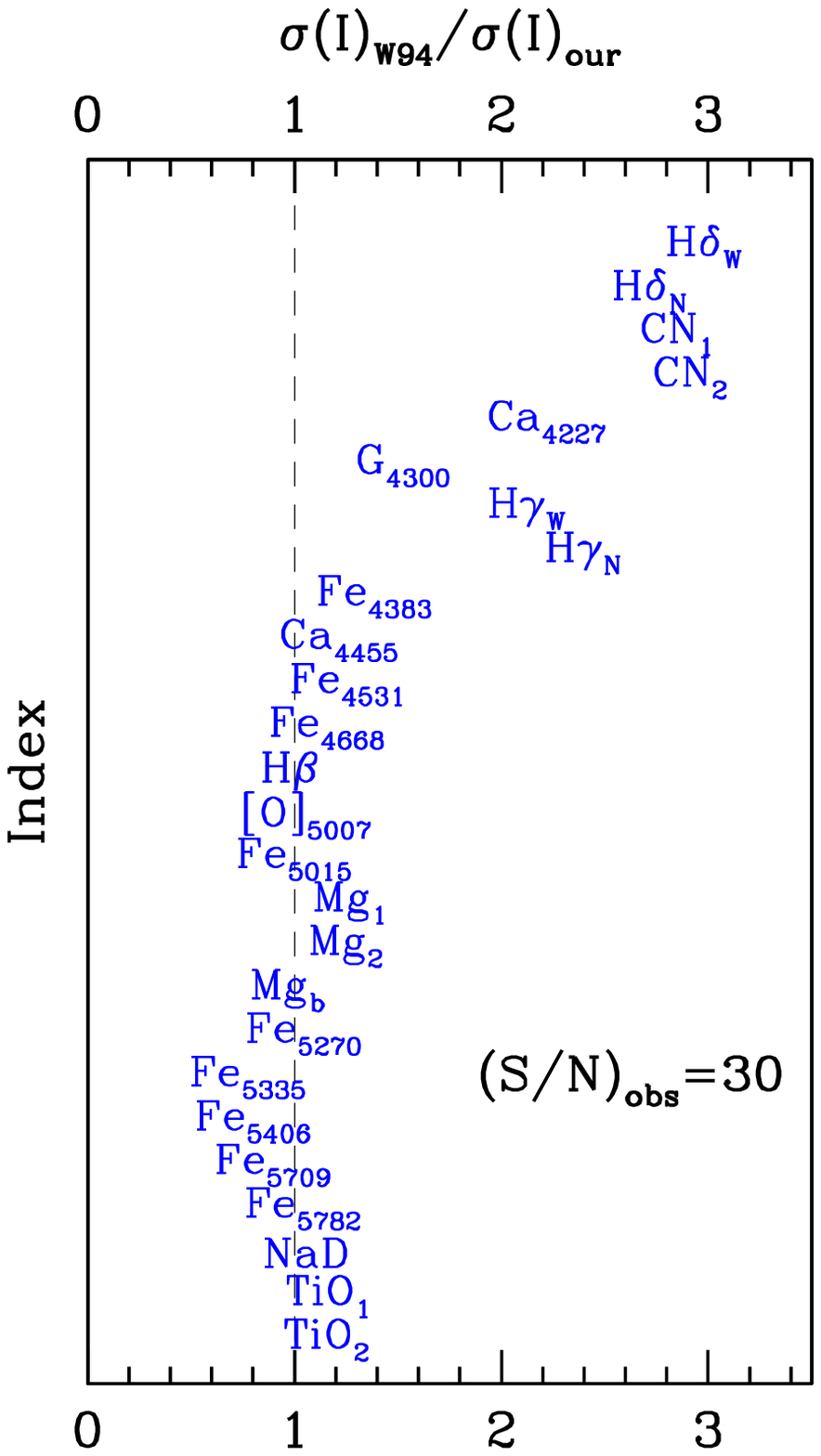}
}
\caption{
The expected index standard deviation $\sigma(I)_{our}$, according to eq.~(\ref{eq:sang}),
for an assumed spectral $S/N \sim 30$ is compared with the empirical estimate from the Lick primary 
calibrators, as from Table~1 of \citet{worthey94} and \citet{worthey97}. The uncertainty for 
[OIII]$_{5007}$ has been estimated from the \citet{gonzalez93} data, as from Table 4.1 therein, by scaling down
his figures to $S/N = 30$. Notice that Lick IDS spectra lack any stable intensity response, 
a feature that prevents any firm assessment of the S/N ratio of the original data \citep{faber76,worthey94b}.
A remarkably good agreement is found, however, with our theoretical predictions, as far as ``red'' indices
(about $\lambda \ga 4350$~\AA) are concerned. A notable worsening in the Lick calibration accuracy 
is evident at shorter wavelength, likely due to poorer detector performance and statistical drawbacks
inherent to index definition, as discussed in the text.
}
\label{f2}
\end{figure}

Although no firm appraisal can be done of original IDS data in terms of S/N ratio,
a problem extensively discussed by \citet{worthey94b}, nonetheless 
there is point to believe that the S/N assumed in the present derivation is similar to the 
floor imposed on the IDS spectra by small-scale flat-fielding problems. In this 
framework one could recognize our statistical analysis as an instructive excercise to 
put in a more standard context the inherent uncertainty of 
Worthey's et al reference dataset, at least as far as indices ``redder'' than 
$\sim 4350$~\AA\ are concerned. 

Note, however, a dramatic worsening of the IDS performance in the 
blue, where index accuracy quickly degrades.
This trend can now easily be understood on the basis of our previous statistical arguments.
At shorter wavelength, in fact, the combined effects of a worse IDS response and an 
intrinsic drop of SED for stars of intermediate and late 
spectral type all conspire in the sense of exacerbating any spectral slope, thus 
departing from the standard Gaussian scenario (in force of eq.~\ref{eq:rg}), and 
adding further extra-volatility to Lick-index calculation.

\section{Index entanglement and redundancy}

A closer look to Fig.~\ref{f1} data shows that about 25\% of the covered wavelength range actually 
contributes to the definition of two or more indices leading, on average, to a ~$1.5$ oversampling 
fraction. Such a redundancy evidently calls for a more detailed analysis of the possible 
bias effects that in some cases make indices inter-dependent in their stochastic fluctuation,
that is {\it beyond any strictly physical relationship.} 

This is certainly the case of some intervening emission, that superposes to absorption 
features ($H\beta$ is a typical example in this sense), 
but also partly or fully overlapping pseudo-continuum windows, in common with different 
indices, may induce some degree of correlation in the corresponding output
(as for the blue sidebands of Ca4455 or Fe5335, fully comprised within the Fe4383 and
Fe5270 red sidebands, respectively). This may also happen to feature windows, as shown 
for instance by Mgb, fully comprised within the Mg$_2$ central band. 
Finally, a further possibility is that the entire feature contributes to 
the pseudo-continuum window of another index: the case of Fe5335, 
is explanatory in this regard being fully nested into the red sideband of Mg$_2$ and Mg$_1$.

In this section we want therefore to assess some illustrative cases giving, when
possible, a guideline for a more general application of our results.
The theoretical predictions that come from the reference discussion of Sec.~2, 
will be accompanied by an empirical check, that could allow us a clean assessment 
of the different situations sketched above.

\subsection{Radial spectroscopy of elliptical galaxies as a robust empirical check}

For our aims it is convenient to rely on a set of spectroscopic data for a sample 
of 20 early-type (mainly elliticals) galaxies observed with the 2.12m telescope of the 
``G. Haro'' Observatory of Cananea (Mexico) in a series of runs along years 1996/97.
These observations are part of a more general exhaustive study on spectroscopic 
gradients across the surface of galaxies and its physical implications 
for galaxy evolution \citep{carrasco95,buzzoni09b,buzzoni15}. It is clearly beyond the scope of 
this work to dig further into the original scientific issue and the detailed 
data reduction analysis (the reader could be addressed to the cited references 
for a full review). 
We just want to recall here the basic information to characterize the sample in view of 
its possible relevance as an independent tool to probe our theoretical output.

The spectra have been collected with a long-slit B\"oller \& Chivens Cassegrain 
spectrograph equipped with a 300 gr/mm grating, that provided a dispersion of 
67 \AA/mm along a 4200-6000 \AA\  wavelength interval. Spectral resolution
was set to 5 \AA\ (FWHM) throughout ($R = \lambda/\Delta \lambda \sim 1000$)
Galaxy targets were typically exposed along one hour ($3\times 1200''$ frames) 
centering along galaxy major axis. The resulting spectra have been wavelength-
and flux-calibrated according to standard procedures, and have then been 
rebinned in the spatial domain such as to sample galaxy radius at 3-5 values
of distance symmetrically placed with respect to the center, roughly up to 
one effective radius \citep[see][for an example]{buzzoni09b}. 
The rebinning procedure allowed us to greatly improve the $S/N$ ratio 
especially in the outermost regions such as to work there with $S/N \ga 25$~px$^{-1}$,
a figure that quickly improved to $S/N \ga 80$~px$^{-1}$ toward the center.

The relevant information of the galaxy sample is summarized in Table~\ref{t2}.
From the RC3 catalog \citep{devaucouleurs91} we report morphological classification,
absolute $B$-band magnitude (M$_B$), and apparent $(B-V)$ color within one effective 
radius ($r_e$). In addition, in the last two columns we added, respectively, the radial fraction 
sampled with our spectra (r/r$_e$) \citep[see][for further details]{buzzoni09b}, and
a flag for galaxies with reported emission. In this regard, one asterisk mark
a confirmed [OII]$_{3727}$ emission, as in the \citet{bettoni87} catalog, while with
a double asterisk we marked targets with evident [OIII]$_{5007}$ and/or H$\beta$ 
central emission in our spectra.

\begin{table}
\caption{The galaxy database}
\label{t2}
\begin{tabular}{llcccl}
\hline\hline
NGC & Type  & M$_B$ & (B-V) & $r_{max}/r_e$ & emission?\\
\hline
1587 & E1p    & -21.19 & 1.04 &  0.2 & * * \\
1588 & Ep     & -19.91 & 1.00 &  0.5 &     \\
2685 & S0-a   & -19.09 & 0.94 &  0.8 & *   \\
2764 & S0     & -19.55 & 0.71 &  1.2 & * * \\
3245 & S0     & -20.09 & 0.89 &  1.0 &     \\
3489 & S0-a   & -19.23 & 0.85 &  0.8 & * * \\
3607 & E-S0   & -20.00 & 0.93 &  0.4 & *   \\
4111 & S0-a   & -19.16 & 0.88 &  3.0 & *   \\
4125 & E6p    & -21.27 & 0.94 &  0.4 & *   \\
4278 & E1     & -19.35 & 0.96 &  1.0 & * * \\
4374 & E1     & -21.05 & 1.00 &  0.6 & *   \\
4382 & S0-a   & -20.50 & 0.89 &  0.3 &     \\
4472 & E2     & -21.71 & 0.98 &  0.8 &     \\
4649 & E2     & -21.47 & 1.00 &  0.5 &     \\
4742 & E4     & -19.35 & 0.79 &  1.3 & *   \\
5576 & E3     & -20.14 & 0.90 &  0.4 &     \\
5812 & E0     & -20.44 & 1.04 &  0.4 &     \\
5866 & S0-a   & -20.00 & 0.93 &  0.7 & * * \\
5982 & E3     & -21.38 & 0.94 &  0.7 & *   \\
6166 & E2p    & -22.96 & 1.03 &  0.4 & *   \\
\hline\hline
\end{tabular}
\end{table}

For the galaxies of Table~\ref{t2}, \citet{buzzoni09b,buzzoni15} have derived the 
complete set of 17 standard Lick indices (including the [OIII]$_{5007}$ feature), 
from G4300 to Fe5709. These data provide us with a useful reference to perform, 
for each index, a series of {\it statistically equivalent realizations}. 
This can be carried out in force of the supposed central symmetry of the galaxy 
spectral properties. Even in case of spectroscopic radial gradients, in fact, 
a similar stellar population has to be hosted at the same distance on opposite 
sides of the galaxy leading to a nominally identical set of absorption indices. 
Clearly, any systematic deviation from this pure folding effect may be worth of 
attention in our analysis.

Note that, as we deal with index {\it differences}, our method is quite insensitive 
to standardization problems (zero points settlement, calibration drifts etc.), 
although we obviously magnify the internal random errors of our analysis by a factor 
of $\sqrt{2}$. About 80 couples of index differences have been computed across the 
full galaxy sample, for a total of roughly 1400 measures for the full set of 17 indices.

\begin{figure}
\centerline{
\includegraphics[width=\hsize,clip=]{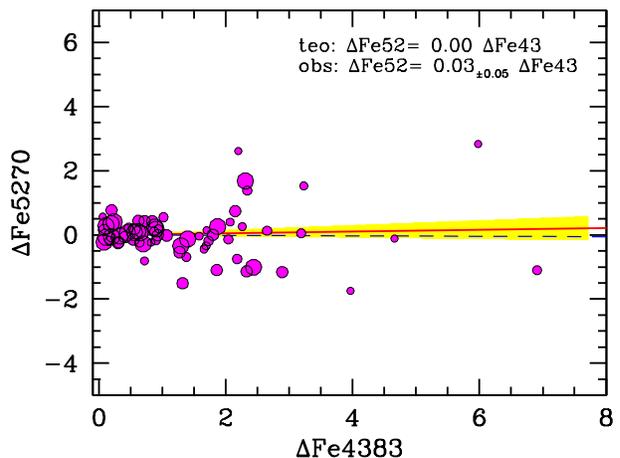}
}
\caption{
The envisaged $\Delta$-$\Delta$ test among galaxy spectroscopic indices in the
\citep{buzzoni09b,buzzoni15} galaxy sample (see text for details)
is explained here for
the illustrative case of Fe4383 (index ``x'') and Fe5270 (index ``y''). 
If we assume the source of ``left -- right'' index 
difference to be only of stochastic nature, then no correlation has to be expected between
$\Delta$~Fe4383 and $\Delta$~Fe5270. 
A least-square linear fit to the data, in the form
of $\Delta y = \alpha \Delta x$, actually provides $\alpha = 0.03_{\pm 0.05}$ (solid line with
the yellow fan marking the $\pm 1 \sigma$ standard deviation of the slope estimator), fully 
consistent with a flat trend (dashed line), as expected for 
this case. Just as a guideline, notice that the dot size in the plot is smaller for index 
measurements at farther radial distances from galaxy center.
}
\label{f3}
\end{figure}

Just as an explanatory example, to illustrate our approach, we show in Fig.~\ref{f3} 
the $\Delta$-$\Delta$ index distribution for two FeI indices, namely Fe5270 versus 
Fe4383. Each point in the plot 
comes from the following procedure. For two slices at the same radial distance (on opposite 
side) of a given galaxy, we computed $\delta x$ as the ``left -- right'' difference of the Fe4383
index measurements. The same is done for Fe5270, deriving $\delta y$. We then set 
$\Delta x = \vert\delta x\vert$ and obtain the sign of $\delta x$ as ${\rm sign}_x = \delta(x)/\Delta(x)$, 
so that $\Delta y = {\rm sign}_x \delta y$. 
Though, from a physical point of view, both indices should correlate with Iron abundance,
no mutual influence can exist for their relative variation. As a consequence, their 
``left-right'' differences must be statistically independent over the entire galaxy sample.
This is actually confirmed by the least-square fit to the data in the figure (68 points
in total), which provides $\alpha = 0.03_{\pm 0.05}$.\footnote{As a general rule, 
throughout in our $\Delta$-$\Delta$ fits, we applied {\it both} to $x$ and $y$ variables 
an rms clipping procedure at a $3\sigma$ level to reject the 
catastrophic outliers (residual spikes from cosmic rays in the spectra, cold/warm pixels 
affecting the index strength, lacking index estimate on either ``left'' or ``right'' side etc.).}

\subsection{{\textit {\textbf M\'enage \`a trois}}: H$\beta$, [OIII]$_{5007}$ and Fe5015}

The wavelength region of galaxy spectra around 5000~\AA\ may be heavily prone to the effects 
of gas emission. In particular, the $H\beta$ and Fe5015 strengths could be affected, 
the first resulting from the parallel action of stellar absorption and gas emission,
the latter because of the possible [OIII] 5007\AA\ appearence (see Fig.~\ref{f4}).
This issue is of central relevance as, within the Lick system, the $H\beta$ index is 
the most suitable one to place strong constraints to the age of unresolved stellar 
aggregates \citep[e.g.][and many others]{buzzoni94,bressan96,vazdekis96,
tantalo98,kobayashi99,maraston00,trager00,thomas05,fusipecci05,sanchez06b,peletier07}
due to its close response to the temperature location of the Main Sequence Turn Off
stars \citep{gorgas93,buzzoni95,buzzoni09a}.

A proper correction of this effect is not quite a straightforward task, 
especially when modest amounts of emission do not appear as ``surging'' features in the
spectra and rather hide making absorption line-depth artificially shallower. If 
uncorrected, this effect would bias age estimates toward older values.
A first attempt to account for $H\beta_e$ residual emission is due to \citet{gonzalez93}, 
who assumed the [OIII]$_{5007}$ index to be a confident proxy for $H\beta$ correction.
A simple scheme was devised in this sense, where the observed Balmer index should be offset 
by $\Delta H\beta = H\beta_e = -0.7 [OIII]_{5007}$ \citep{gonzalez93}. This approach relies on
the fact that [OIII] emission would be much more clearly recognizable than possibly 
hidden $H\beta$ emission, and [OIII] and Balmer emission do somewhat correlate as they
ostensibly originate in the same regions of the galaxies. 
One has to note, however, that later attempts to settle the problem empirically 
\citep{carrasco95,trager00,kuntschner06} reached puzzling conclusions, at odds with Gonzalez' 
predictions.

\begin{figure}
\centerline{
\includegraphics[width=\hsize,clip=]{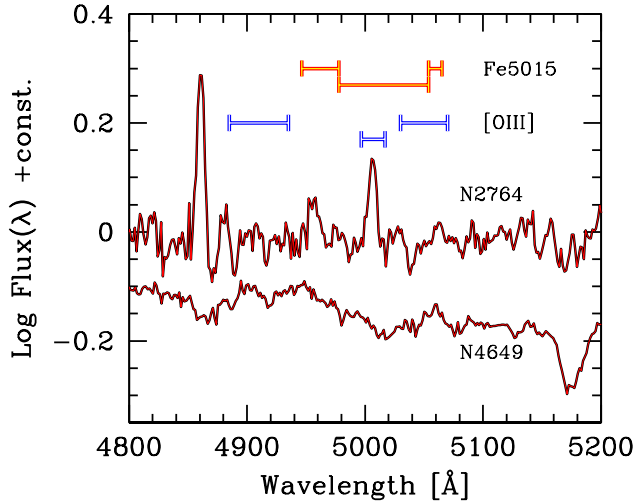}
}
\caption{
The nested configuration of the [OIII]$_{5007}$ and Fe5015 indices. For each index the
two blue and red pseudo-continuum sideband are reported, flanking the main
feature window. Two representative spectra of active (NGC~2764) and quiescent (NGC~4649) 
elliptical galaxies, from our observed database \citep{buzzoni09b}, are superposed to the index sketch, 
for the sake of comparison. In addition to the [OIII] emission, note for NGC~2764 
also a prominent H$\beta$ emission at the nominal $\lambda\lambda\, 4861$~\AA.
}
\label{f4}
\end{figure}

To a closer analysis, things may actually be much more entangled as {\it i)} apparent 
[OIII] emission is partly at odds with Fe5015 absorption, as both indices partly
overlap (see Fig.~\ref{f4}); 
{\it ii)} fresh star formation could likely trigger both [OIII] and
Balmer emission but on different timescales \citep{moustakas06,buzzoni15}, and depending on Oxygen
abundance and ionization parameters \citep{osterbrock74,pagel79,kewley02}. For all these reasons, 
the combined behaviour of the three indices, H$\beta$, [OIII]$_{5007}$ and Fe5015, 
needs to be assessed self-consistently.

\begin{figure}
\centerline{
\includegraphics[width=\hsize,clip=]{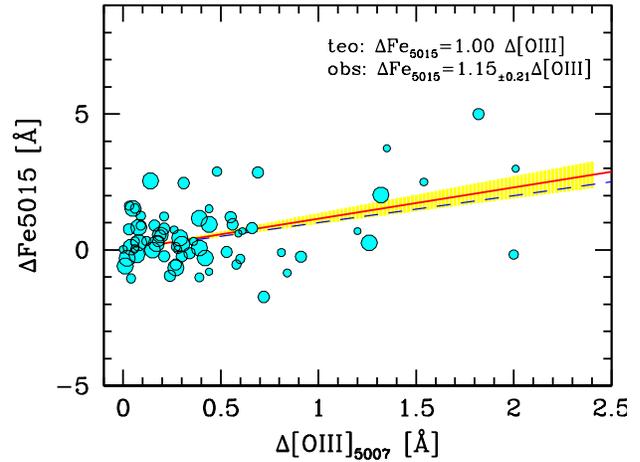}
}
\caption{
The $\Delta$-$\Delta$ correlation test, as for Fig.~\ref{f3}, for the [OIII]$_{5007}$
and Fe5015 indices. The expected statistical prediction, according to eq.~(\ref{eq:f50o50})
(dashed line), is well matched by the observations, the latter ones providing a fully consistent 
$\alpha = 1.15_{\pm 0.21}$ fitting line slope (solid line within a $\pm 1 \sigma$ yellow fan), 
as reported top right in the plot.
}
\label{f5}
\end{figure}

It is useful to start our analysis first with the [OIII]$_{5007}$ versus 
Fe5015 interaction. Figure~\ref{f4} clearly shows that Oxygen emission may be systematically
underestimated as the [OIII]$_{5007}$ continuum sidebands cannot account for Fe5015 
absorption. This is especially true if one considers that Fe5015 strengthens in old stellar 
systems \citep{trager00,beuing02,ogando08,buzzoni15} (see, for instance, the illustrative 
case of galaxy NGC~4649, also displayed in Fig.~\ref{f4}).
Let us then compute the observed flux density within the 
[OIII] and Fe5015 feature windows (of width $w_O$ and $w_{50}$, respectively).

By recalling eq.~(\ref{eq:ia}) and (\ref{eq:r1}), and omitting the ``average'' overlined notation for
flux density symbols, as well as the suffix to the Oxygen index, for better legibility of the 
formulae, we have
\begin{equation}
{f_O\over f_c} = \big(1-{[OIII] \over w_O}\big), \qquad
{f_{50}\over f_c} = \big(1-{Fe5015 \over w_{50}}\big).
\label{eq:22}
\end{equation}
Note that $f_O$ is in fact a sum of the genuine contribution of
O emission plus the contribution of the {\it intrinsic} Fe5015 feature.
Similarly, the intrinsic Fe5015 flux density (i.e.\ after fully removing O emission) is
\begin{equation}
{f'_{50}\over f_c} = \big(1-{Fe5015' \over w_{50}}\big),
\label{eq:23}
\end{equation}
where Fe5015' stands for the corrected Fe5015 index strength.
By definition, it must be
\begin{equation}
f'_{50}(w_{50} -w_O) = f_{50}\, w_{50} - f_O\,w_O
\label{eq:24}
\end{equation}
so that, after matching eq.~(\ref{eq:22}),
\begin{equation}
{f'_{50}\over f_c} = \Big[ {{w_{50} \big(1-{Fe5015 \over w_{50}}\big) - w_O \big(1-{[OIII] \over w_O}\big)} \over {(w_{50}-w_O)}}\Big].
\label{eq:25}
\end{equation}
When combining eq.~(\ref{eq:23}) and (\ref{eq:25}) this lead to 
\begin{equation}
Fe5015' = w_{50}\ \big( 1- {f'_{50}\over f_c} \big) = {w_{50} \over {w_{50} - w_O}}\ \big(Fe5015 - [OIII]\big).
\label{eq:26}
\end{equation}
By replacing the corresponding width of each feature window (namely $w_O = 20$~\AA\ and
$w_{50} = 76.3$~\AA, see Table~\ref{t1}), the intrinsic Fe5015 index can eventually be written 
explicitely in terms of the observed quantities as 
\begin{equation}
Fe5015' = 1.36\ \big(Fe5015 - [OIII]\big).
\label{eq:27}
\end{equation}

As far as the $\Delta$-$\Delta$ index distribution is concerned, if the same (intrinsic) 
value of Fe5015' has to be expected in the galaxy ``left'' and ``right'' side, at fixed 
radial distances, then any difference in the observed Fe5015 index may be induced by 
a change in the [OIII] emission. By simply differentiating eq.~(\ref{eq:27}), we obtain 
a straight positive correlation such as
\begin{equation}
\Delta Fe5015 = \Delta [OIII].
\label{eq:f50o50}
\end{equation}
Again, when comparing in Fig.~\ref{f5} with our data, a least-square solution on
68 selected points in total (after a $3\sigma$ clipping) provides a fully consistent 
slope value $\alpha = 1.15_{\pm 0.21}$.

Quite importantly, turning back for a moment to eq.~(\ref{eq:27}), one has to notice 
that, when [OIII] emission fades and Fe5015 tends to Fe5015', then the
[OIII] index {\it does not vanish}. In fact, if we set $Fe5015 = Fe5015'$ in 
eq.~(\ref{eq:26}), one obtains
\begin{equation}
[OIII]_{\rm min} = {w_O \over w_{50}} Fe5015' = 0.26\ Fe5015'
\label{eq:28}
\end{equation}
More generally, if some emission occurs, then $[OIII]' = [OIII]-[OIII]_{\rm min}$, and
\begin{equation}
[OIII]' = [OIII]-0.26\ Fe5015',
\label{eq:29}
\end{equation}
or
\begin{equation}
[OIII]' = 1.36\ [OIII]-0.36\ Fe5015
\label{eq:30}
\end{equation}
putting it in terms of the observed quantities.
Equation~(\ref{eq:29}) gives the locus in the Fe5015 versus [OIII] plane
for any intrinsic emission strength $[OIII]'$. 

By combining eq.~(\ref{eq:26}) and (\ref{eq:29}), an obvious self-consistency 
constraint implies that 
\begin{equation}
Fe5015' = Fe5015 - [OIII]'.
\label{eq:32}
\end{equation}

\begin{figure}
\centerline{
\includegraphics[width=\hsize,clip=]{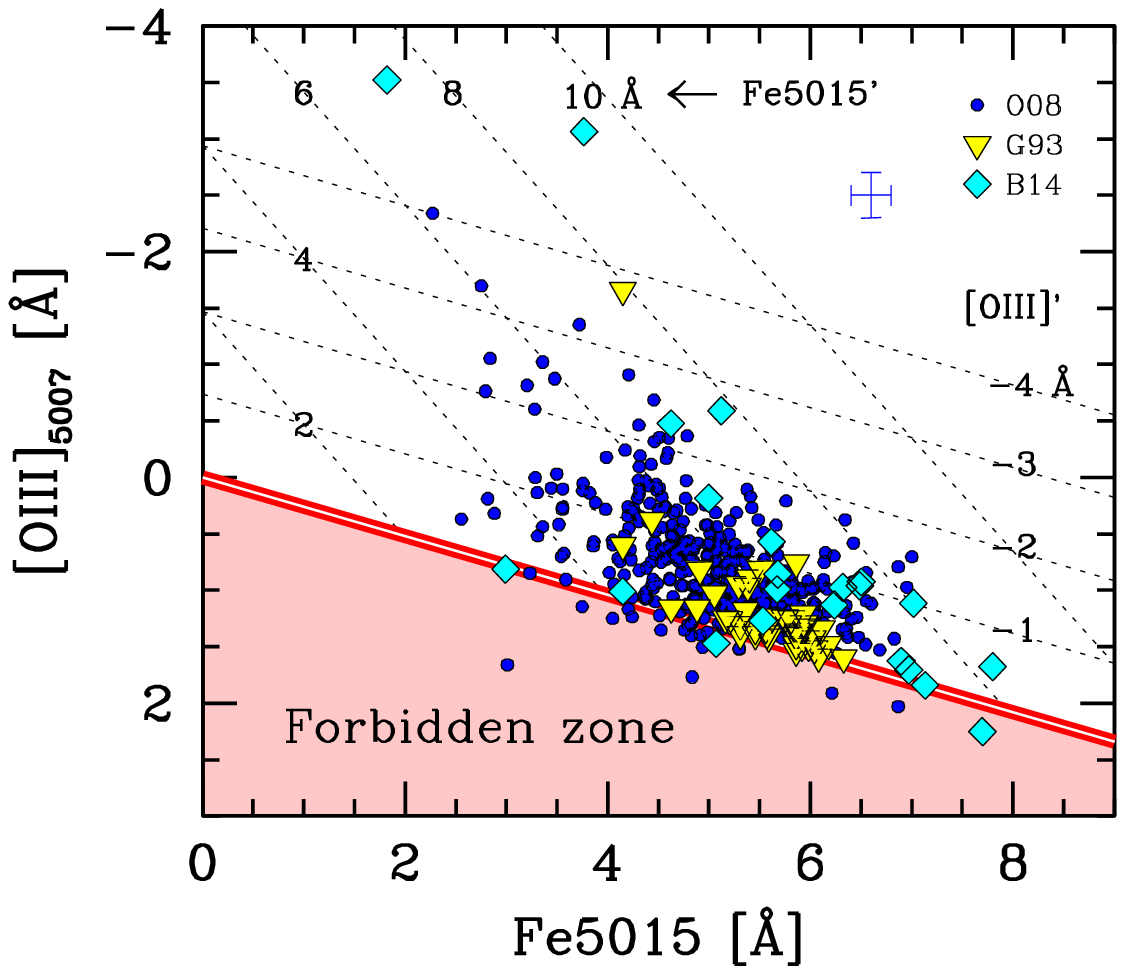}
}
\caption{
The observed (raw) [OIII]$_{5007}$ versus Fe5015 index distribution for the
\citet{ogando08} (solid dots), \citet{gonzalez93} (triangles) and \citep{buzzoni09b,buzzoni15}
(rombs) samples of early-type galaxies. The forbidden zone in the index domain
is shaded, according to eq.~(\ref{eq:30}), while a more general parameterization
versus intrinsic Fe5015 and [OIII]$_{5007}$ strength (primed indices in the plot) 
is displayed, in force of eq.~(\ref{eq:27}) and (\ref{eq:30}).
An indicative error box for the observations is reported top right.
}
\label{f6}
\end{figure}

A graphical summary of our theoretical scheme is sketched in Fig.~\ref{f6}, where we compare
the [OIII] versus Fe5015 distribution for the galaxies of Table~\ref{t2} with the observed 
distribution of an extended sample of 509 ellipticals from \citet{ogando08} and with the 
\citet*{gonzalez93} original bulk of 41 ellipticals. The trend seems quite encouraging, 
and actually calls for an important property of ellipticals.
Within the limits of \cite*{ogando08} calibration (see in particular Table 7 and 8 therein),
in fact, these data indicate that a substantial fraction of galaxies displays some 
residual Oxygen emission \citep{sarzi06,papaderos13}.
This important issue can be better assessed by means of Fig.~\ref{f7}, where we derive the
distribution of the intrinsic emission $[OIII]'$ as from eq.~(\ref{eq:30})
The whole sample of 498 galaxies with available Fe5015 and [OIII]$_{5007}$ indices leads to a 
mean index strength $\langle{[OIII]'}\rangle = -0.70\pm0.54$~\AA, with an evident skewness 
toward enhanced [OIII] emissions. Just as a reference, 16 objects out of 498 (3\%) in the
figure display an intrinsic [OIII] feature stronger than 2~\AA, while for 107 galaxies (21\%)
the [OIII] emission exceeds 1~\AA.

\begin{figure}
\centerline{
\includegraphics[width=\hsize,clip=]{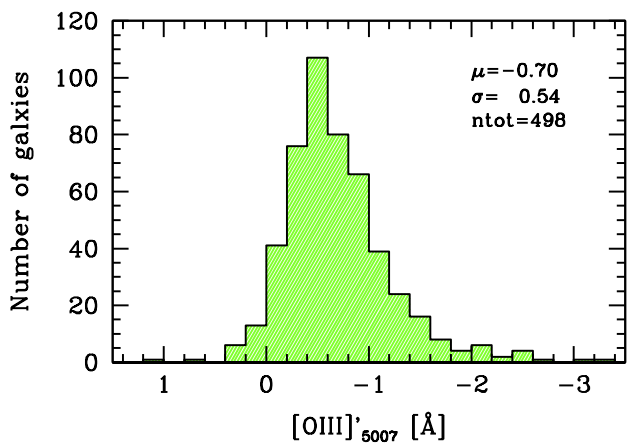}
}
\caption{
The [OIII]$_{5007}$ distribution of the intrinsic (``primed'') index among
the \citet{ogando08} early-type galaxy sample. A total of 498 galaxies, 
with available [OIII]$_{5007}$ and Fe5015 observations are displayed.
The [OIII] observations have been corrected according to eq.~(\ref{eq:30}).
On average, ellipticals display a weak residual gas emission with
$\langle [OIII]'_{5007}\rangle = -0.70_{\pm 0.54}$~\AA. For a fraction of 21\%
of the total sample, the intrinsic [OIII] pseudo-equivalent width is in excess 
of 1~\AA, while about 1/7 of these galaxies (16 out of 498) display a striking 
(star-formation) activity, with [OIII]'$_{5007} \la -2$~\AA.
}
\label{f7}
\end{figure}

After disentangling the combined behaviour of [OIII] emission and Fe5015 absorption,
we are now in a better position to tune up an H$\beta$ correction scheme.
Again, our $\Delta$-$\Delta$ diagnostic plot may add useful hints to tackle the problem.
The rationale, in this case, is that the observed $\Delta [OIII]_{5007}$ should directly
trace the corresponding [OIII]' intrinsic change, via eq.~(\ref{eq:29}), providing
the Fe5015' strength is the same at symmetric distances from the galaxy center.
We could therefore probe the $\Delta H\beta$ data set for any possible correlation
with the corresponding  $\Delta [OIII]_{5007}$ output in a fully empirical way.

For this task, we divided the whole galaxy set of Table \ref{t2} (Sample ``A'') in different 
subsamples according to the ``activity level'' as traced by the overall spectral characteristics.
In particular, we considered three refence samples. Sample ``B'' consists of the 7 ``quiescent''
galaxies with no reported emission lines. Sample ``C'' includes, on the contrary,
the 8 ``moderately active'' objects, marked with one asterisk in the table, wich are 
reportedly emitting 
at least the [OII] 3727~\AA\ feature. Finally, the 5 ``most active'' ellipticals (that 
is those marked with double asterisk in Table~\ref{t2}) are included in Sample ``D''. 
They display [OII] 3727~\AA\ emission accompanied by the strongest [OIII]' index in 
Fig.~\ref{f6}. Our results are summarized in the four panels of Fig.~\ref{f8}.
As usual, the fitting slope after $3\sigma$ data clipping is presented 
in Table~\ref{t5} together with the mean of central corrected [OIII] emission,
\citep[as from][]{buzzoni15}.

\begin{figure}
\includegraphics[width=0.83\hsize,clip=]{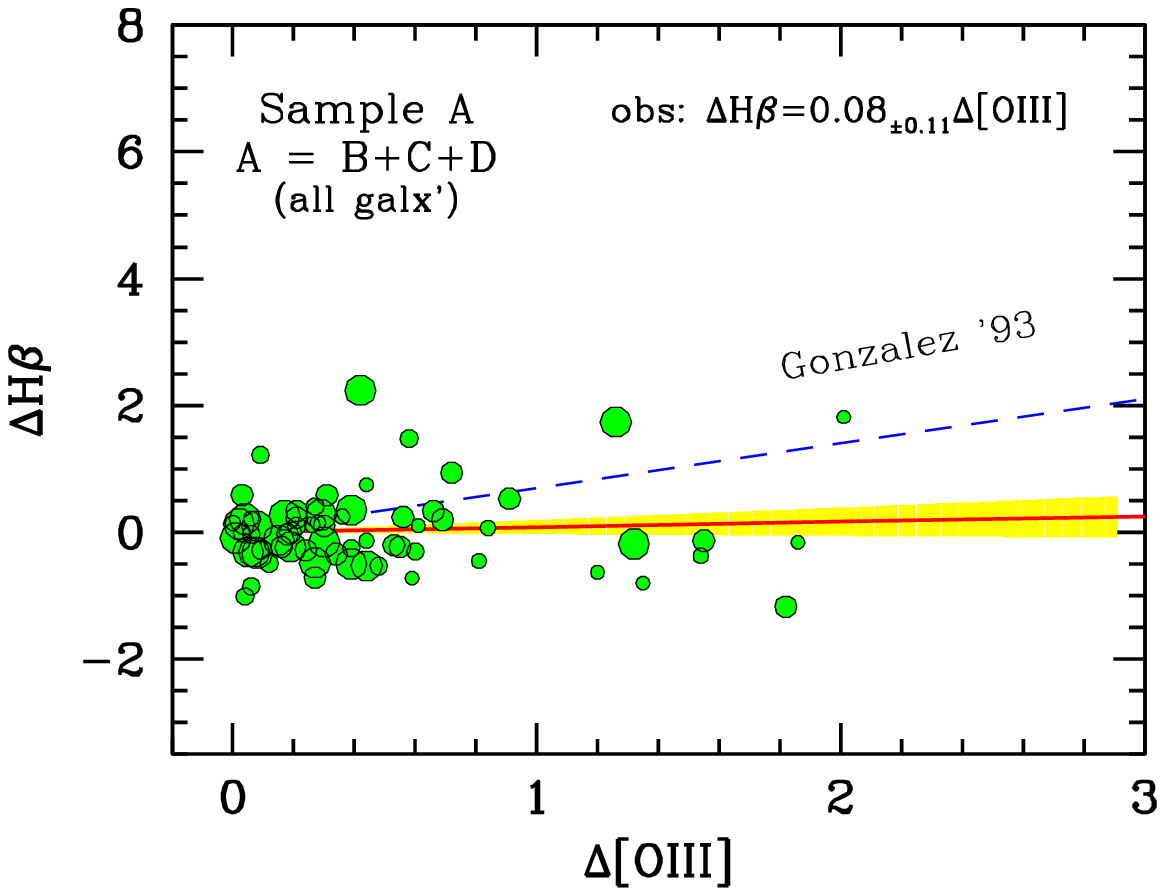}
\includegraphics[width=0.83\hsize,clip=]{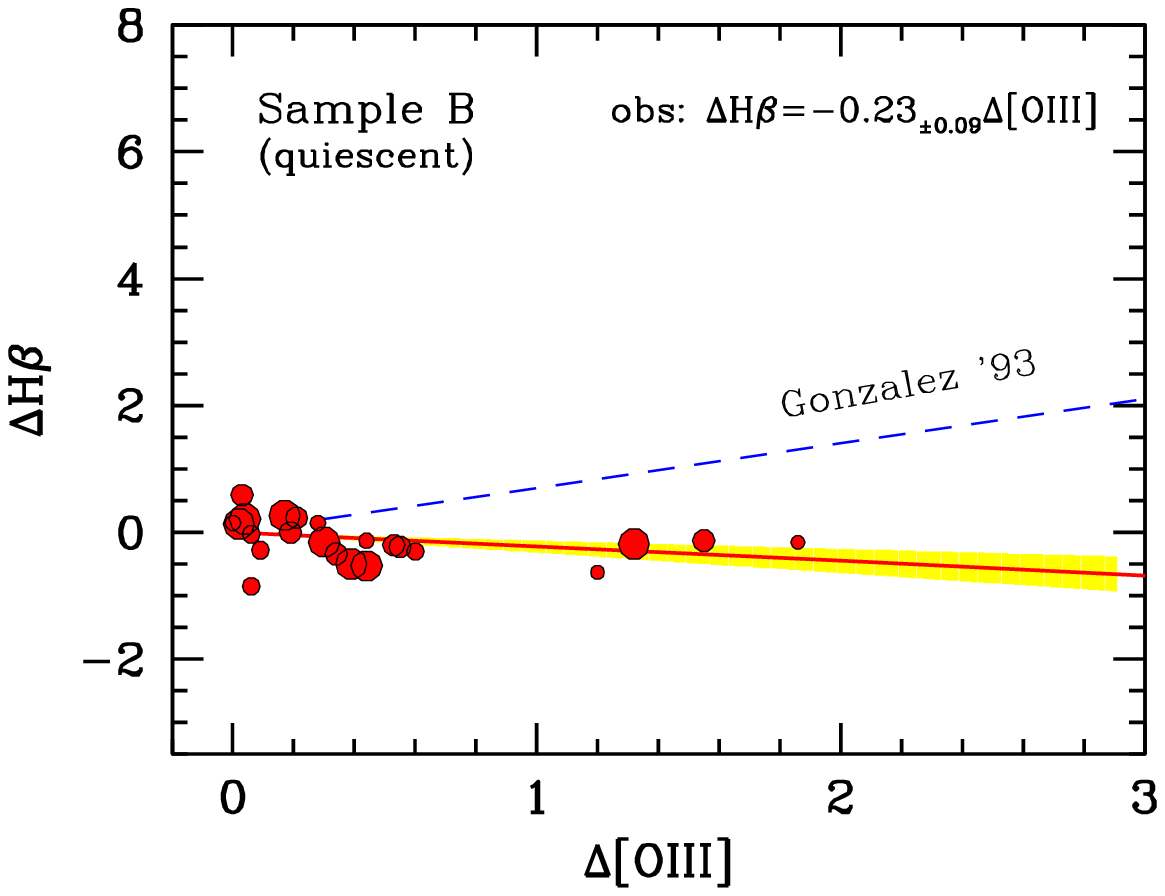}
\includegraphics[width=0.83\hsize,clip=]{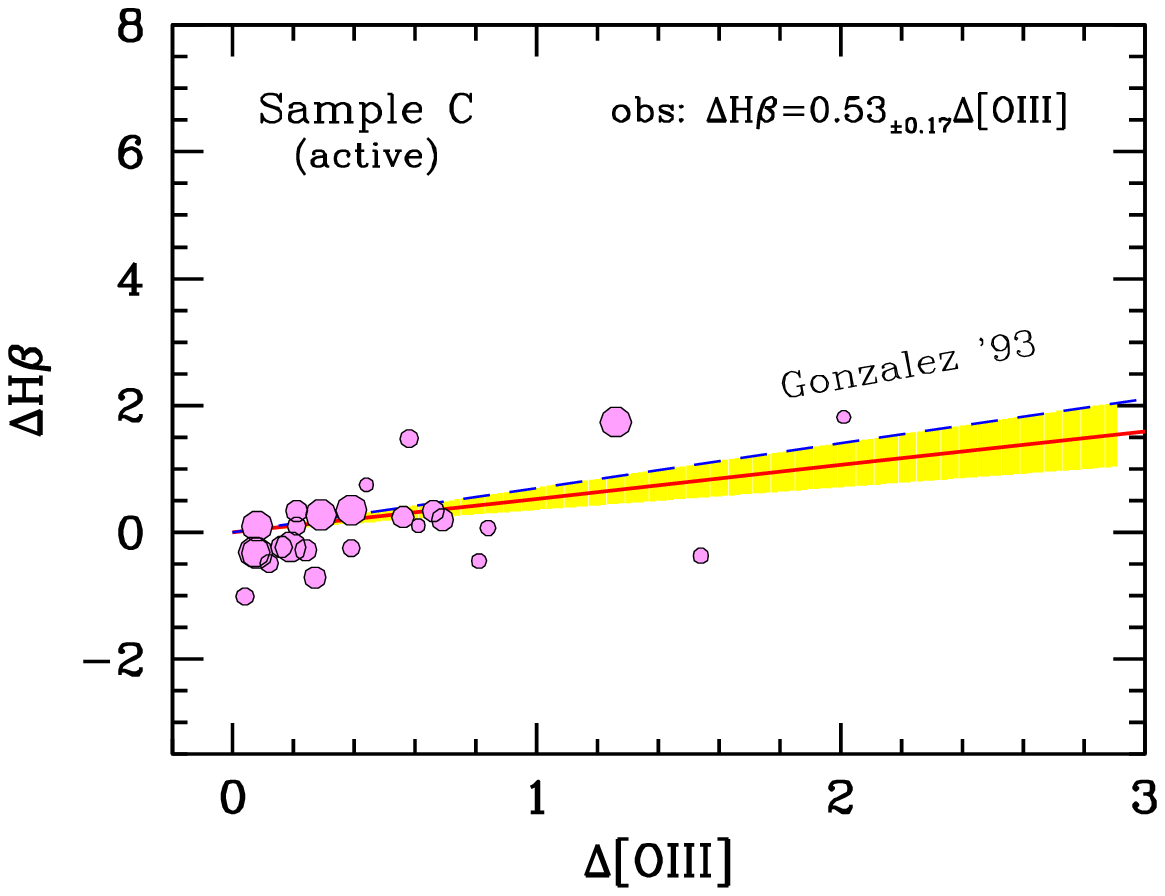}
\includegraphics[width=0.83\hsize,clip=]{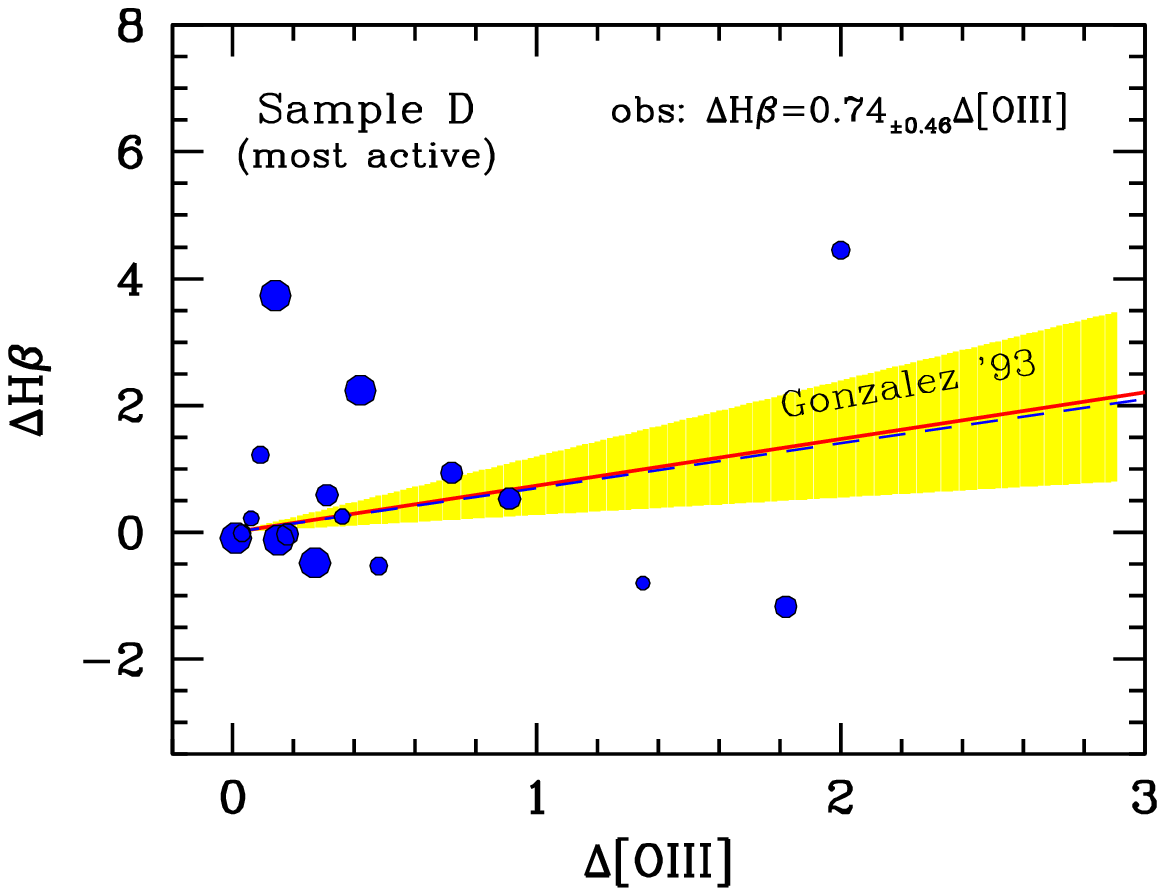}
\caption{
The $\Delta$-$\Delta$ correlation test, as for Fig.~\ref{f3}, for the [OIII]$_{5007}$
and H$\beta$ indices in the adopted \citet{buzzoni09b,buzzoni15} sample of early-type galaxies.
The whole Sample A, of the 20 objects (upper panel), is split into three groups
(i.e.\ Samples ``B'', ``C'' and ``D'', from top to bottom) consisting, respectively of
7, 8, and 5 galaxies with increasing gas emission activity. Groups have been assembled according 
to the asterisk ranking of Table~\ref{t1}, that combines the \citet{bettoni87} 
empirical classification with the [OIII] index strength in our observations.
The dashed line, in each plot, marks the \citet{gonzalez93} $\Delta$-$\Delta$ correction slope.
Our least-square fit is reported top right in each panel, and displayed (solid line) together 
with its corresponding $\pm 1 \sigma$ uncertainty (yellow fan). Our output is summarized in Table~\ref{t5}.
}
\label{f8}
\end{figure}

\begin{table}
\caption{Core [OIII]' emission strength and observed $\Delta H\beta$ versus $\Delta [OIII]$ fitting slope}
\label{t5}
\begin{tabular}{llrc}
\hline\hline
Sample & \multicolumn{1}{c}{$\langle$[OIII]'$\rangle^\dagger$} & $\alpha$ & n$^\ddagger$ \\
\hline
A - All galaxies (20 obj)   &  --1.1         	 &  0.08& (68)\\
                            & {\tiny $\pm 1.7$}  & {\tiny $\pm 11$} & \\
B - Quiescent only (7 obj)  & --0.15		 & --0.23& (23)\\
                            & {\tiny $~\pm 74$} & {\tiny $\pm 9$} &  \\
C - Interm. active (8 obj)  & --0.67		 &  0.53& (25)\\
                            & {\tiny $~\pm 49$} & {\tiny $\pm 17$} & \\
D - Most active (5 obj)     & --3.1	         & 0.74& (17)\\
                            & {\tiny $\pm 2.3$}  & {\tiny $\pm 46$} & \\
\hline\hline
\noalign{$^\dagger$ Mean core [OIII]' index, in \AA, as from \citet{buzzoni15}}
\noalign{$^\ddagger$ Number of available pairs of points in the $\alpha$ fit}
\end{tabular}
\end{table}

A firm important conclusion of our analysis, is that no striking correlation seems in place
across the whole galaxy sample between H$\beta$ and [OIII]$_{5007}$.
A mild positive slope is found overall (Sample ``A'', upper left panel in Fig.~\ref{f8}), only 
marginally in excess to the flat-slope case ($\alpha = 0.0$-0.2). The flattening effect is 
clearly induced by the fraction of ``quiescent'' ellipticals (Sample ``B'', upper right panel), 
for which we even notice a slight anti-correlation ($\alpha \sim -0.2$). However, with increasing 
stellar activity, 
as in Sample ``C'' and ``D'' galaxies, one has to notice some hints of a change. Although with 
increased scatter and a much blurred trend, the data seem to point to a steeper slope 
$\alpha$ and, at least for Sample ``D'' (lower left panel in the figure), our results 
nominally match \cite*{gonzalez93} value.

As a plain ``thumb rule'', we could state 
that \cite*{gonzalez93} correction scheme fully holds for the most active strong-lined 
ellipticals, where $[OIII]'$ emission roughly exceeds 2~\AA.
For intermediate cases (say for [OIII]' emission strength about 1~\AA) a flatter slope
($\alpha \sim 0.3$-0.5) might be more appropriate, while a negligible H$\beta$ correction 
($\alpha \sim 0.1$) is required for weaker [OIII] emission. As shown in Fig.~\ref{f9},
a logistic curve could be adopted as a convenient analytical fit of this scheme, such as
\begin{equation}
\alpha = {{0.75}\over{1+e^k}},
\label{eq:33a}
\end{equation}
with $k = ([OIII]'+1)/0.25$, and [OIII] emission assumes, of course, negative values.
The corrected H$\beta$ index therefore derives as
\begin{equation}
H\beta_{\rm corr} = H\beta_{\rm obs} -H\beta_e = H\beta_{\rm obs} -\alpha\,[OIII]'.
\label{eq:33}
\end{equation}

\begin{figure}
\centerline{
\includegraphics[width=\hsize,clip=]{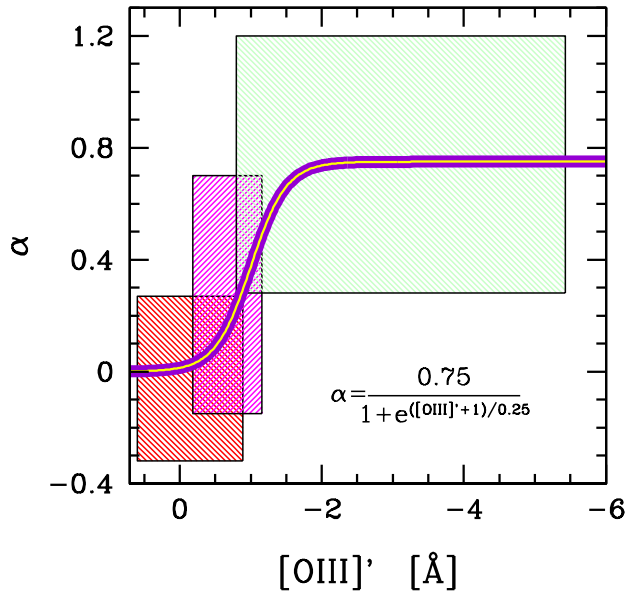}
}
\caption{
The devised correction scheme for H$\beta$ in the form of a logistic curve versus intrinsic
[OIII]' emission, as from eq.~(\ref{eq:33a}). The ``quiescent'', ``active'' and ``most active'' 
galaxy samples of Fig.~\ref{f8} are sketched by the three shaded boxes, with [OIII]' mean 
central values and $\alpha$ error bars as from Table~\ref{t5}.
}
\label{f9}
\end{figure}

\subsection{Mg$_1$ + Mgb = Mg$_2$?}

The prominent Magnesium feature about 5200~\AA\ is by far the most
popular target in optical spectroscopy of elliptical galaxies and other old stellar systems.
As explored in full detail by \citet{mould78}, from a theoretical point of view, this feature
is in fact a blend of both the atomic contribution of the Mgb triplet at 5178~\AA, and 
the (1,1) $A^2\Pi\ X^2\Sigma^+$ vibrational band head of the MgH  molecule.
This transition of Magnesium Hydride has a dissociation energy of 1.34~eV 
(\citealp{coelho05}, as from \citealp{huber79} data), a much lower threshold than the 
first-ionization potential of 7.65~eV for atomic Mg. As a consequence, the Mgb absorption
tends to be the prevailing one in warmer (type F-G) stars, while it accompanies the 
surging MgH absorption at the much cooler temperature regime of K-M stars.
In addition, the relatively more flimsy structure of the molecule makes MgH very
sensitive to stellar luminosity class \citep{buzzoni01}, as gravity acts on 
electron pressure of the stellar plasma and slightly modulate the dissociation energy threshold.
This dependence, however, is severely tackled by the intervening appearence of
TiO bands among M-type stars, which substantially affect the pseudo-continuum level
\citep[][see, for instance Fig. 5 therein]{tantalo04,buzzoni92}.

\begin{figure}
\centerline{
\includegraphics[width=\hsize,clip=]{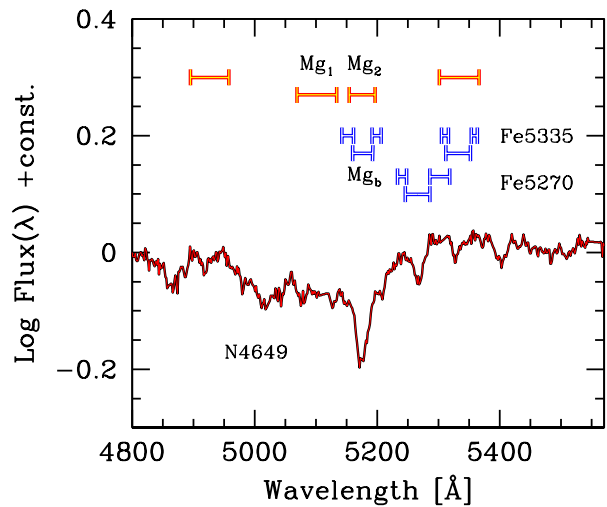}
}
\caption{
An illustrative sketch of the Mg index set, and the two Fe5270 and Fe5335 indices.
For each index, the blue and red sidebands are marked, flanking the relevant feature window. 
The nested configuration of the Mgb, Fe5335 and Mg$_2$ indices is well evident, as discussed
in the text. The representative spectrum of the elliptical galaxy NGC~4649 \citep[from the][data]{buzzoni09b} 
is superposed for the sake of comparison.
}
\label{f10}
\end{figure}

The Magnesium complex is sampled by three Lick indices, namely Mg$_1$, Mg$_2$, and Mgb. 
The strategy is for Mg$_2$ to somewhat bridge the other two, with Mg$_1$ especially suited 
to probe the molecular contribution, and the Mgb index better tuned on the atomic triplet
(see Fig.~\ref{f10}). Within this configuration, Mg$_2$ shares its 
feature window with Mgb, and its pseudo-continuum windows with the  Mg$_1$ index. From its 
side, Mgb aims at recovering the genuine atomic contribution by setting its pseudo-continuum 
level about the Mg$_1$ bottom (see again Fig.~\ref{f10} for an immediate sketch).

Similarly to what discussed in previous section, we can further expand our scheme  
to assess in a more straightforward way the Mg$_2$-Mgb-Mg$_1$ index entanglement.
First of all, as both Mg$_2$ and Mg$_1$ are expressed in magnitudes and Mgb in \AA,
it is convenient to recall the basic relations of eq.~(\ref{eq:imag}) and (\ref{eq:4}) and write

\begin{figure}
\includegraphics[width=\hsize,clip=]{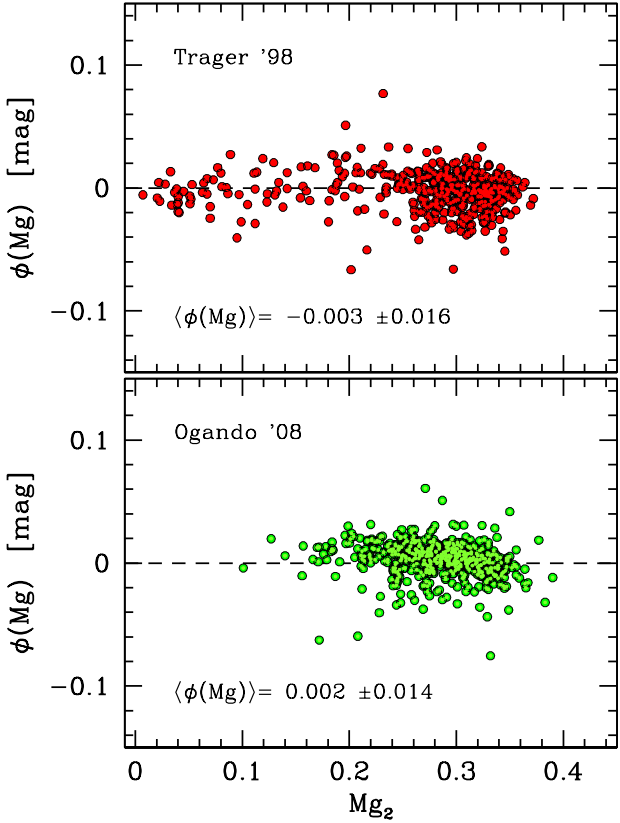}
\caption{
An empirical check of the expected relationship among the three popular Mg indices sampled by the Lick system.
The synthesis function $\phi(Mg)$ of eq.~(\ref{eq:38}) has been scrutinized by means of the two extended
early-type galaxy samples of \citet{trager98} (406 galaxies with complete Mg$_1$, Mg$_2$ and Mgb data, 
see upper panel), and \citet{ogando08} (481 galaxies, lower panel). The match with both samples allowed
us to obtain the fine-tuning offset in the theoretical relationship, such as $k = 0.03$~mag.
After offset application, notice for both plots the lack of any drift in the point distribution, a 
feature that assures the correct parameterization of our theoretical relationship. In addition, the 
point scatter around the zero line indicates that index mutual relationship can be obtained within
a $\sim$1.5\% internal accuracy.
}
\label{f11}
\end{figure}

\begin{eqnarray}
\left\{
\begin{array}{lcl}
{f_2 / f_c} &=& 10^{-0.4\,\mathrm{Mg}_2} \\
\\
{f_1 / f_c} &=& 10^{-0.4\,\mathrm{Mg}_1} \\
\\
{f_b / f_1} &=& \left(1 -{{\rm Mgb}/ w_b}\right).
\end{array}
\right.
\label{eq:34}
\end{eqnarray}
We can also combine the last two equations above to obtain
\begin{equation}
{f_b \over f_c} = \left({f_b \over f_1}\right) \left({f_1 \over f_c}\right) = \left(1 -{{\rm Mgb}\over w_b}\right)\, 10^{-0.4\,{\rm Mg}_1}.
\label{eq:34b}
\end{equation} 
Quantities $f_2, f_1$, and $f_b$ in previous equations refer to the flux density of Mg$_2$, Mg$_1$ 
and Mgb feature windows, respectively, and $w_b$ is the width of the Mgb feature window. For Mg$_1$ and 
Mg$_2$ this flux is normalized to the same pseudo-continuum, f$_c$, while the f$_1$ bottom level 
can also be seen as a proxy of the Mgb pseudo-continuum. 

Likewise eq.~(\ref{eq:24}), also for the Mg case we can write
\begin{equation}
f_2\,w_2 = f_1\,(w_2-w_b) +\left({f_b\over f_1}\right)\,f_1\,w_b,
\label{eq:35}
\end{equation}
being $w_2 = 42.5$ \AA\ and $w_b = 32.5$ \AA\ (from Table~\ref{t1}) the width of the Mg$_2$ and Mgb 
feature windows, respectively.
By matching eq.~(\ref{eq:34}), (\ref{eq:34b}) and (\ref{eq:35}) 
and taking the logarithm, after little arithmetic, we lead to the final form that links the
three indices:
\begin{equation}
{\rm Mg}_2 - {\rm Mg}_1 = -2.5\,\log \left(1 -{{\rm Mgb}\over w_2}\right) +k.
\label{eq:mg}
\end{equation}

In the equation we left a fine-tuning offset ($k$) in order to account for the little 
aproximation we made in eq.~(\ref{eq:34}) for f$_1$ to be an exact proxy of the Mgb pseudo-continuum 
level. It is interesting to verify, on empirical basis, the invariancy of the function
\begin{equation}
\phi({\rm Mg}) = {\rm Mg}_2 - {\rm Mg}_1 +2.5\,\log \left(1 -{{\rm Mgb}\over w_2}\right) -k.
\label{eq:38}
\end{equation}
Again, the \citet{ogando08} together with the \citet{trager98} galaxy database provide
us with an excellent reference tool. Our results are summarized in the two panels of
Fig.~\ref{f11}. As predicted, the lack of any drift of $\phi({\rm Mg})$ versus Mg$_2$ (and
versus Mg$_1$ and Mgb, as well) indicates that no residual dependency has been left
unaccounted in our analysis. By searching the least-square fit to the data, we also get
an empirical estimate for the fine-tuning offset, that is $k = 0.03$~mag. 
Providing to observe two indices, eq.~(\ref{eq:mg}) allows us to secure
the third one within a $\sim 1.5$\% internal accuracy.\footnote{Note that, for vanishing values of Mgb
as in low-metallicity stellar populations, eq.~(\ref{eq:mg}) further simplifies as its r.h. term can 
be usefully aproximated by the linear term of its series expansion, i.e.\ $[k+1.09\,({\rm Mgb}/w_2)] \sim 0.03\,({\rm Mgb}+1)$. In this case, Mg 
indices straightforwardly relate as ${\rm Mg}_2 \simeq {\rm Mg}_1 +0.03\,({\rm Mgb}+1)$.}

A further check can be carried out in terms of our $\Delta$-$\Delta$ test. By differentiating
eq.~(\ref{eq:mg}), and considering that, in general, Mgb~$\ll w_2$, we obtain, in fact
\begin{equation}
d{\rm Mg}_2-d{\rm Mg}_1 \simeq {1\over w_2}\,{d{\rm Mgb}\over (1-{\rm Mgb}/w_2)} \approx 0.024\, d{\rm Mgb},
\label{eq:dmg}
\end{equation}
The thoretical relationship is probed in Fig.~\ref{f12} with our galaxy sample of Table~\ref{t2}. 
The usual least-square fit procedure to a total of 66 points provides a value of the slope 
$\alpha = 0.019_{\pm 0.003}$.

\begin{figure}
\includegraphics[width=\hsize,clip=]{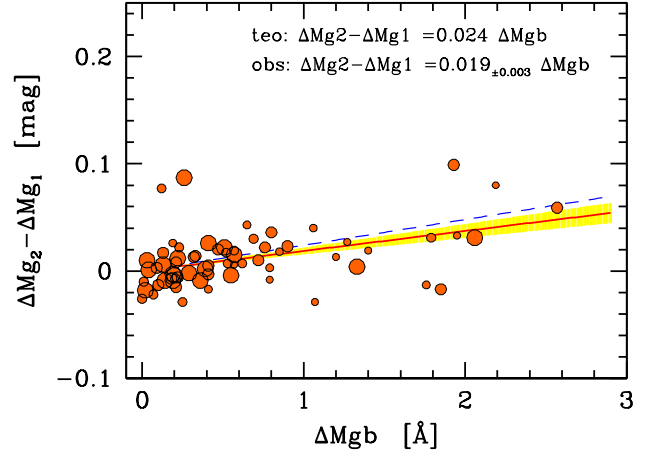}
\caption{
The $\Delta$-$\Delta$ correlation test, as for Fig.~\ref{f3}, for the Mg index complex.
The $\Delta Mg_1$, $\Delta Mg_2$ and $\Delta$ Mgb variation across the \citet{buzzoni09b,buzzoni15} galaxy sample
of Table~\ref{t2} is probed in terms of the expected relationsip, as from eq.(\ref{eq:dmg}) (dashed line
in the plot).
Our least-square fit is reported top right, and displayed (solid line) together 
with its corresponding $\pm 1 \sigma$ statistical uncertainty (yellow fan).
}
\label{f12}
\end{figure}

\subsubsection {The [NI]$_{5199}$ hidden bias}

As for the case of the [OIII]$_{5007}$ perturbing emission to the Fe5015 index, also for the Mgb 
index one may expect a shifty, and mostly neglected, bias from overlapping gaseous emission lines. 
This is the case of the [NI] forbidden doublet at 5198/5200~\AA\ (hereafter [NI]$_{5199}$), a somewhat 
ominous weak feature that characterizes the spectra of LINER galaxies 
\citep[][see Appendix 4 therein]{osterbrock74}, including ellipticals, either due to residual star 
formation activity \citep{sarzi06,cappellari11}, or simply supplied by the post-AGB stellar 
component \citep{kaler80,ferland12,kehrig12,papaderos13}.

\citet{goudfrooij96} have been discussing in some detail the perturbing effect of 
[NI]$_{5199}$ on the computed Mgb index in the spectra of elliptical galaxies. Differently from the
[OIII]$_{5007}$ versus Fe5015 interplay, the [NI]$_{5199}$ emission {\it positively} correlates with the 
affected Mgb as [NI] luminosity enhances the pseudo-continuum level in the red side band of Mgb
(see Fig.~\ref{f10}), thus leading to a stronger index.

\begin{figure}
\includegraphics[width=\hsize,clip=]{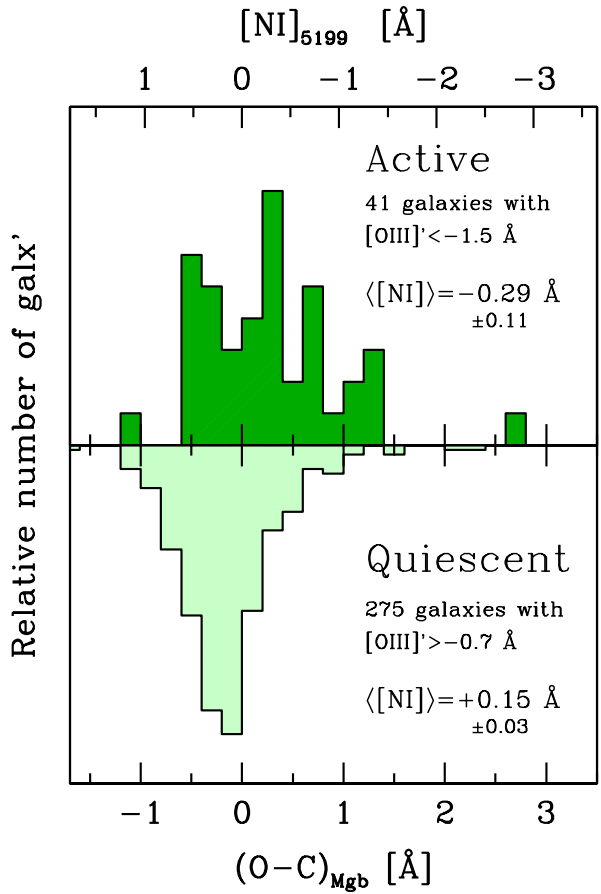}
\caption{
The inferred distribution of the ``hidden'' [NI]$_{5199}$ gas emission in the \citet{ogando08} galaxy
sample. In case of intervening [NI] emission, the Mgb index is expected to enhance, on average, by the amount
of eq.~(\ref{eq:n3}) with respect to its theoretical figure, as from eq.~(\ref{eq:mg}). 
This excess (in the sense ``Observed -- Computed'', see lower x axis), and the related [NI] strength 
(upper x axis) is probed by contrasting the 
subsample of 41 ``active'' galaxies (as traced by explicit [OIII] emission 
stronger than 1.5~\AA, as in the upper histogram) with the population of 275 ``quiescent'' galaxies 
with vanishing [OIII] (see the lower histogram). See text for a discussion.
}
\label{f12b}
\end{figure}

A simple correction scheme can be devised, to size up the [NI]$_{5199}$ bias just relying on
the third relation of eq.~(\ref{eq:34}) (and, again, assuming $f_1$ as the reference Mgb 
pseudo-continuum flux density). By passing to logarithms we have: 
\begin{equation}
\ln\left(1-{{\rm Mgb}\over w_b}\right) = \ln(f_b)-\ln(f_1)
\label{eq:n1}
\end{equation}
Again, considering that Mgb~$\ll w_b$, the l.h.\ term of the equation can be approximated by $-({\rm Mgb}/w_b)$
so that, by differentiating with respect to f$_1$, we have
\begin{equation}
{d{\rm Mgb}\over w_b} \simeq {df_1\over f_1}.
\label{eq:n2}
\end{equation}
If [NI]$_{5199}$ emission appears, then a supplementary flux F$_N$ is added to the Mgb pseudo-continuum,
and its reference flux density changes by $df_1 = F_N/W_b$, where $W_b = 33.7$~\AA, as from Table~\ref{t1}.
As, by definition, the [NI] equivalent width\footnote{Note that we maintain a {\it negative} notation 
for {\it emission} lines. This explains the minus sign in the relevant equations throughout our 
discussion.} is $[NI]_{5199} \simeq -{F_N /f_1}$, 
eq.~(\ref{eq:n2}) can eventually be re-arranged in its final form
\begin{equation}
{\rm dMgb} = -[NI]_{5199} \left(w_b \over W_b\right) \simeq -0.96\, [NI]_{5199}.
\label{eq:n3}
\end{equation}
Our result nicely compares with \citet{goudfrooij96}, who led to a more crude analytical estimate 
of the effect such as ${\rm dMgb} \simeq -1.1\, [NI]_{5199}$.

The \citet{ogando08} extended galaxy database allows us a further independennt assessment of the 
[NI]$_{5199}$ bias among elliptical galaxies. In this regard, we set up a simple test aimed at
comparing the observed Mgb index for the entire galaxy dataset with the corresponding ``predicted'' 
output as from eq.~(\ref{eq:mg}). Galaxies with intervening [NI] 
emission are expected to display (on average) a stronger-than-predicted Mgb index, and one 
could envisage to detect this signature in the index (O-C) (i.e.\ ``observed - computed'') residual
distribution.

In order to get a cleaner piece of information, we only restrained our analysis to the galaxy
sub-sample with stronger and poorer [OIII]$_{5007}$ emission, assuming [OIII]$_{5007}$ 
to be also a confident proxy for [NI]$_{5199}$ activity \citep{osterbrock74}. According to the 
[OIII]'$_{5007}$ distribution of Fig.~\ref{f7}, we therefore selected a sample of 41 ``active'' 
galaxies with [OIII]'$_{5007} < -1.5$~\AA\ emission, to be compared with a group of 275 objects
with average or less-than-average (i.e.\ [OIII]'$_{5007} > -0.7$~\AA) emission.
For each galaxy with observed Mg$_2$ and Mg$_1$ indices we then computed the corresponding Mgb, 
via eq.~(\ref{eq:mg}) (by setting $k=0.03$ therein), and compared it with the observed entry in the
catalog. The output distribution for the ``active'' and ``quiescent'' galaxy samples
is displayed in Fig.~\ref{f12b}. According to eq.~(\ref{eq:n3}), the (O-C) residuals have been directly 
converted into a fiducial [NI]$_{5199}$ emission strength, as in the upper x-axis scale of the
figure. Within the uncertainty of our theoretical procedure, just a glance to Fig.~\ref{f12b} 
confirms the lack of any [NI] emission among ``quiescent'' 
galaxies ($\langle [NI]_{5199}\rangle = +0.15${\scriptsize $\pm 0.03$}~\AA, on average),
while a weak but significant ($>3\sigma$) activity has to be reported, on the contrary for ``active'' 
systems, with $\langle [NI]_{5199}\rangle = -0.29${\scriptsize $\pm 0.11$}~\AA. Interestingly enough, 
the [NI]/[OIII]~$\sim 0.3/1.5$ observed ratio for the ``active'' sub-sample matches in the right 
sense the corresponding empirical figure for the LINER galaxies, as suggested by 
\citet[][see Table~1 therein]{goudfrooij96}.

\subsection{Spectral duality: the case of Fe5335}

Due to wavelength oversampling, along the Lick-index sequence it may even happen that  
individual spectral features are accounted twice being also included in some pseudo-continuum windows
of other indices. One relevant case, in this sense, is the Fe5335 index, which is fully 
comprised inside the (red) pseudo-continuum window of both the Mg$_2$ and Mg$_1$ indices.
In addition, it also entirely shares its own red pseudo-continuum with the blue side-window of Fe5270
(see, again, Fig.~\ref{f10}).
Evidently, this makes both the Fe5270 and the Mg indices to somewhat correlate with
the Fe5335 strength. To explore this effect let us start first with the simpler case of 
Fe5335-Fe5270 entanglement.

According to Table~\ref{t1}, the Fe5270 pseudo-continuum is sampled along a total of 47.5~\AA\
(i.e. red+blue side-windows, $W_{52}$), whose 11.3~\AA\ are shared with the Fe5335 red side-band
($W_{53}^{\rm red}$). The total side-windows for Fe5335 
(i.e.\ $W_{53} = W_{53}^{\rm red} + W_{53}^{\rm blue}$) amount to 21.2~\AA, and
both features are sampled within similar wavelength windows ($w_{52}$ and $w_{53}$)
of 40~\AA\ each. By recalling eq.~(\ref{eq:ia}) and passing to logarithm, for Fe5270 we have
\begin{equation}
\ln\left(1-{Fe5270\over w_{52}}\right) = \ln(f_{52})-\ln(f_{52}^c).
\label{eq:40}
\end{equation}

\begin{figure}
\centerline{
\includegraphics[width=\hsize,clip=]{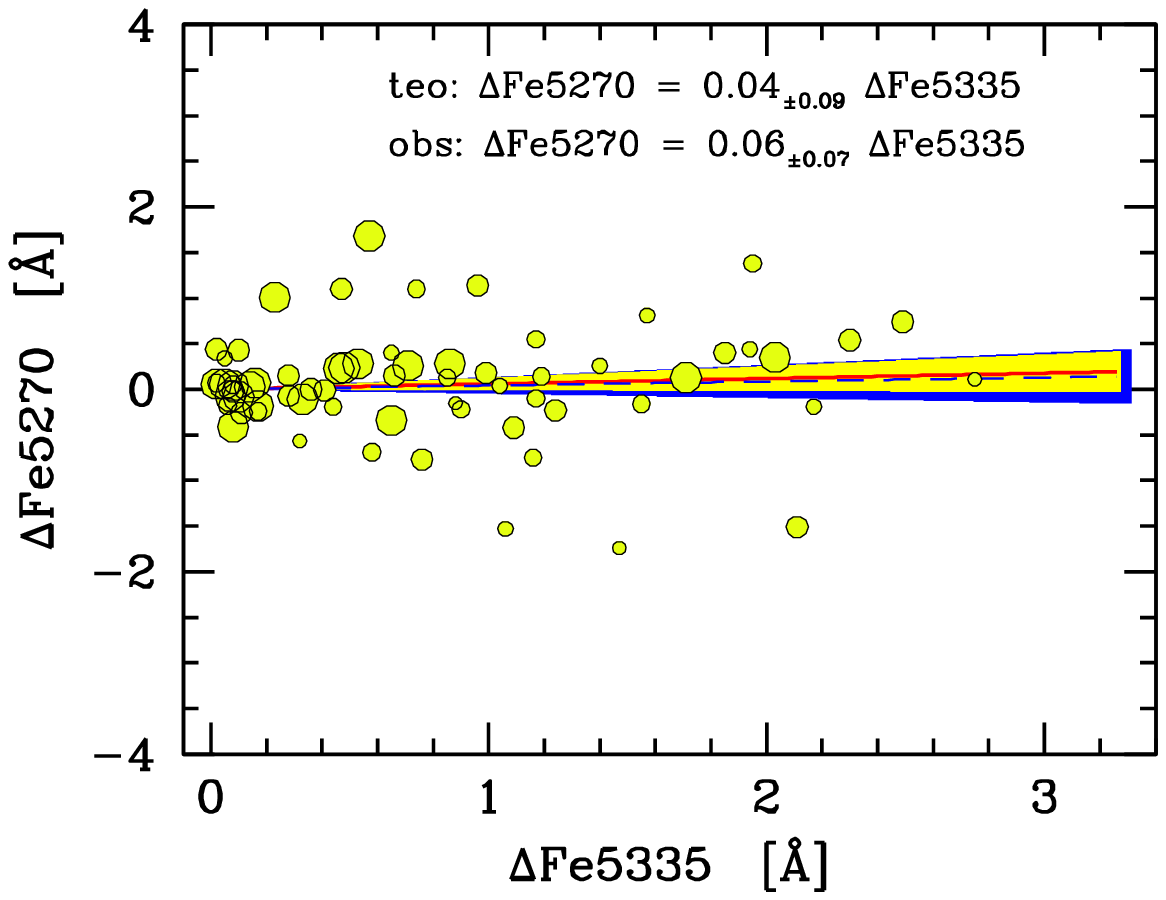}}
\centerline{
\includegraphics[width=\hsize,clip=]{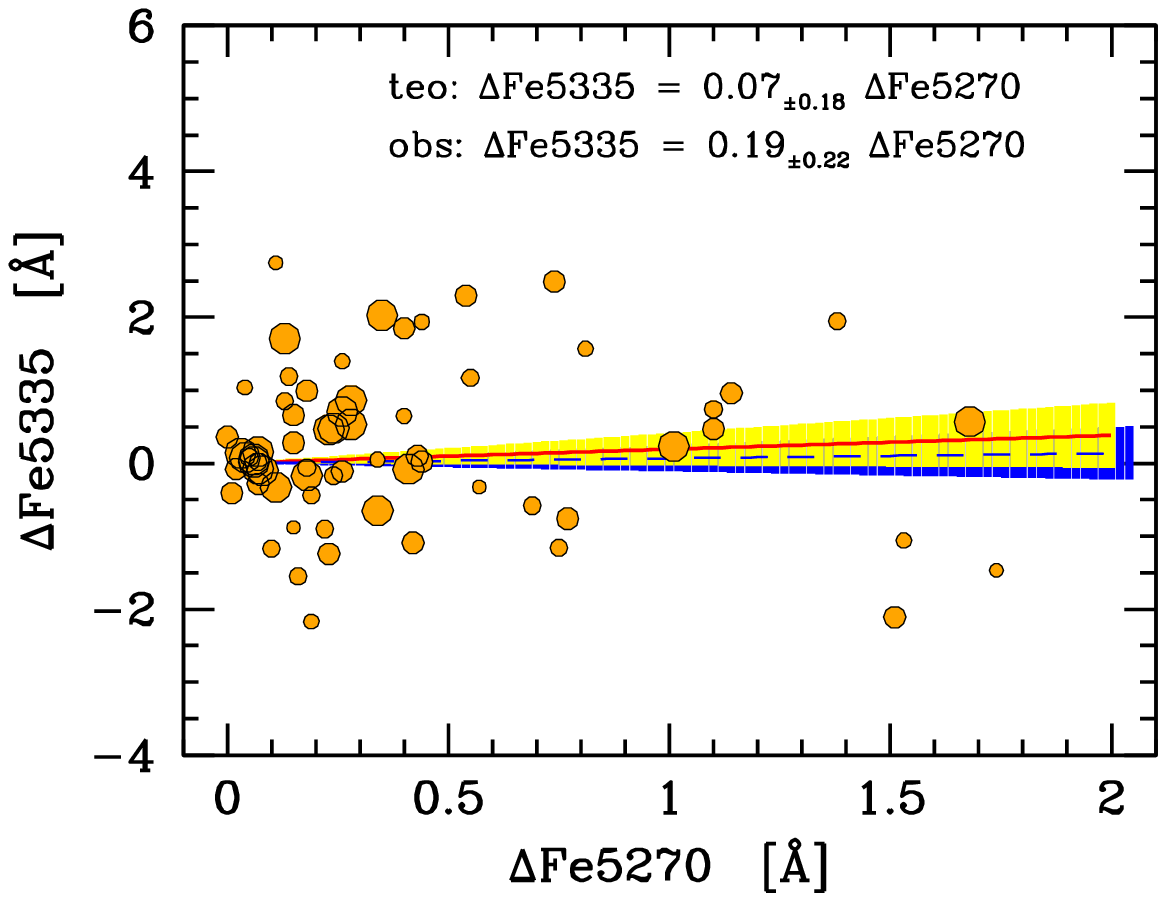}
}
\caption{
The $\Delta$-$\Delta$ correlation test, as for Fig.~\ref{f3}, for the Fe5270 and Fe5335 indices.
The low correlation between index variation, as predicted by eq.~(\ref{eq:44}) and (\ref{eq:46})
(dashed lines in the two plots, with their corresponding $\pm 1 \sigma$ Bayesian uncertainties,
as from the blue fans) is consistently resumed by our observing test on the 
galaxy sample of Table~\ref{t2} (solid lines with their corresponding $\pm 1 \sigma$ statistical
uncertainties, as from the yellow fans). See text for a discussion.
}
\label{f13}
\end{figure}

As we are dealing with a weak absorption line (so that $Fe5270/w_{52} \ll 1$) the l.h. term
of the equation can be approximated by $-(Fe5270/w_{52})$. By differentiating, 
\begin{equation}
\big({dFe5270\over w_{52}})= {df_{52}^c\over f_{52}^c} -{df_{52}\over f_{52}}.
\label{eq:41}
\end{equation}
A similar notation holds for Fe5335, too. Any change (for whatever reason) in the $f_{53}^c$ flux 
density, would therefore reverberate into a change in the Fe5270 and Fe5335 strengths as well, so that 
\begin{equation}
\big({dFe5270\over w_{52}}\big) = {df_{52}^c\over f_{52}^c} \sim {df_{53}^c \over f_{53}^c}{W_{53}^{\rm red}\over W_{52}}
\sim \big({dFe5335\over w_{53}}\big)\big({W_{53}^{\rm red}\over W_{52}}\big),
\label{eq:42}
\end{equation}
or
\begin{equation}
dFe5270 = \big({11.3\over 47.5}\big)\,dFe5335 = 0.238\, dFe5335.
\end{equation}
The chance for Fe5335 to entangle Fe5270 depends on the probability for any random change in 
the Fe5335 index bands to affect the wavelength range in common with the Fe5270 
pseudo-continuum. This can be be estimated of the order 
of $W_{53}^{\rm red}/(W_{53}+w_{53}) \sim 11.3/61.2 = 18$\% being, therefore, 82\% the
probability for no entanglment at all between the two indices.
By accounting for these figures, the {\it average} expected relationship between Fe5270 and 
Fe5335 eventually results in
\begin{equation}
\begin{array}{rl}
dFe5270 &= (0.238\times 0.18) + (0\times 0.82)\, dFe5335 \\
        &= 0.043_{\pm 0.091}\,dFe5335.\\
\end{array}
\label{eq:44}
\end{equation}
This value has to be compared with the $\Delta$-$\Delta$ test as displayed in the upper panel of
Fig.~\ref{f13}. From our data (66 points in total) we obtain $\alpha = 0.059_{\pm 0.068}$.

Quite importantly, note that eq.~(\ref{eq:44}) cannot be simply reversed, once considering the
impact of the overlapping window on the Fe5335 strength (because $W_{53} \neq W_{52}$).
By reversing the indices in eq.~(\ref{eq:42}), in fact we have
\begin{equation}
\big({dFe5335\over w_{53}}\big) = {df_{53}^c\over f_{53}^c} \sim {df_{52}^c \over f_{52}^c}{W_{53}^{\rm red}\over W_{53}}
\sim \big({dFe5270\over w_{52}}\big)\big({W_{53}^{\rm red}\over W_{53}}\big),
\label{eq:45}
\end{equation}
or
\begin{equation}
dFe5335 = \big({11.3\over 21.2}\big)\,dFe5270 = 0.533\, dFe5270.
\end{equation}

This interaction has a probability of the order of $W_{53}^{\rm red}/(W_{52}+w_{52}) \sim 11.3/87.5 = 13$\%.
so that, similar to eq.~(\ref{eq:44}), the eventual relationship, by reversing 
the indices in a $\Delta$-$\Delta$ test becomes
\begin{equation}
\begin{array}{rl}
dFe5335 &= (0.533\times 0.13) + (0\times 0.87)\, dFe5270 \\
        &= 0.07_{\pm 0.18}\,dFe5270.
\end{array}
\label{eq:46}
\end{equation}
Again, this relationship can be probed by means of our $\Delta$-$\Delta$ test, as shown in the
lower panel of Fig.~\ref{f13}, suggesting $\alpha = 0.19_{\pm 0.22}$.

As far as the Mg$_2$- (or Mg$_1$-) Fe5335 entanglement is concerned, we can proceed in
a similar way, depending whether a change occurs in the feature or in the side-band
windows of the FeI index.
According to Table~\ref{t1}, the Mg$_2$ (or Mg$_1$) pseudo-continuum is sampled along 
a total of $W_2 = 127.5$~\AA. Of these, 40~\AA\ and 21.2~\AA\ are shared, 
respectively, with the feature ($w_{53}$) and the side bands ($W_{53}$) of the 
Fe5335 index.

By recalling that Mg$_2$ (and Mg$_1$) are expressed in magnitude, the equivalent
of previous eq.~(\ref{eq:42}), in case of an intervening change in the Fe5335
pseudo-continuum flux, becomes 
\begin{equation}
d{\rm Mg}_2 \sim {df_{2}^c\over f_{2}^c} \sim {df_{53}^c \over f_{53}^c}{W_{53}\over W_2}
\sim \big({dFe5335\over w_{53}}\big)\big({W_{53}\over W_2}\big),
\label{eq:47}
\end{equation}
which leads to
\begin{equation}
d{\rm Mg}_2 = \big({21.2\over 40 \times 127.5}\big)\,dFe5335 = +0.0042\, dFe5335.\\
\label{eq:48}
\end{equation}
This outcome has a probability of the order of $W_{53}/(W_{53}+w_{53}) \sim 21.2/61.2 = 35$\%.

On the other hand, any change in the Fe5335 feature window will affect the Mg$_2$ (Mg$_1$)
strength by
\begin{equation}
d{\rm Mg}_2 \sim {df_{2}^c\over f_{2}^c} \sim {df_{53} \over f_{53}}{w_{53}\over W_2}
\sim -\big({dFe5335\over w_{53}}\big)\big({w_{53}\over W_2}\big).
\label{eq:49}
\end{equation}
By replacing the relevant quantities, in this case we have
\begin{equation}
d{\rm Mg}_2 = -\big({1\over 127.5}\big)\,dFe5335 = -0.0078\, dFe5335,
\label{eq:50}
\end{equation}
with a probability of $w_{53}/(W_{53}+w_{53}) \sim 40/61.2 = 65$\%.
Therefore, if a random process (of whatever origin) is at work by changing Fe5335, then the weighted 
average of eq.~(\ref{eq:48}) and (\ref{eq:50}) eventually provides us with the
expected relationship for Mg$_2$ (and similarly, for Mg$_1$, as well):
\begin{equation}
dMg_2 = dMg_1 = -0.0036_{\pm 0.0057}\, dFe5335.\\
\label{eq:51}
\end{equation}
Our theoretical output is probed in Fig.~\ref{f14}, by averaging the $\Delta$Mg$_1$ and $\Delta$Mg$_2$
versus $\Delta$Fe5335  variations for a cleaner display of the effect. Observations 
actually confirm the nearly vanishing relationship in place across the whole galaxy sample, making 
the expected impact of Fe5335 random change on Mg indices negligible. 

\begin{figure}
\centerline{
\includegraphics[width=\hsize,clip=]{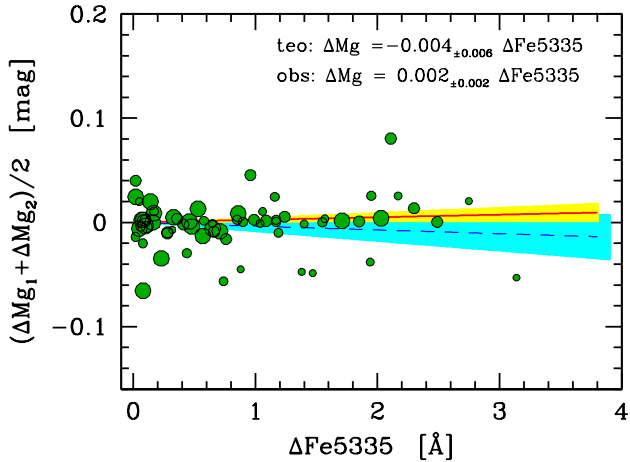}
}
\caption{
The $\Delta$-$\Delta$ correlation test, as for Fig.~\ref{f3}, for the Fe5335 and Mg indices.
The observed relationship for Mg$_1$ and Mg$_2$ is averaged, to gain a cleaner view of the
effect. In spite of the {\it average} low correlation, as predicted by eq.~(\ref{eq:51}) 
(dashed line) one has to notice the important Bayesian uncertainty (blue fan in the plot, 
tracing the $\pm 1 \sigma$ uncertainty), which makes the bias effect of Mg-Fe5335 entanglement 
possibly of non-negligible impact on individual galaxies.
}
\label{f14}
\end{figure}

On the other hand, if a change in the Fe5335 feature occurs as a result of a genuine (selective) 
variation of Fe abundance, then only eq.~(\ref{eq:50}) has to be taken as a reference. Although 
small, the induced bias on Mg$_2$ (and Mg$_1$) may be an issue in this case, expecially if we 
aim at carefully investigating the [$\alpha$/Fe] ratio in a galaxy sample. In fact, any stronger Fe5335 
index would induce, by itself, a shallower Mg index.

\subsection{Mr. Iron, I suppose...}

About half of the total Lick indices are meant to trace Iron as a main elemental
contributor to the index feature. 
The ominous presence of weak (i.e.\ non-saturated) absorption lines of FeI and FeII
in the optical/UV wavelength region is, of course, a well established property of stellar 
spectra \citep[e.g.][]{pagel97}, that can be exploited by high-res spectroscopy to 
set up the linear branch of the spectral curve of growth, thus leading to an accurate 
measure of the elemental abundance. This figure is usually taken as an empirical proxy of 
stellar ``metallicity'', at large, assuming Iron to scale according to all the other metals.
On the same line, also UBV broad-band photometry may rely on the ``blanketing'' effect,
mainly driven by the Iron contribution, to constrain stellar metallicity  
\citep[e.g.][]{sandage69,golay74}.

As far as low-res spectroscopy or narrow-band photometry is concerned, as for the Lick 
indices, however, such a huge number of low-strength Fe lines may reveal a quite tricky 
feature, as Iron markers actually come heavily blended with other elements.
The potential drawback has already been emphasized, at least in a qualitative way, in 
earlier works dealing with the Lick-system assessment 
\citep[][among others]{burstein84,faber85,gorgas93,worthey94b}.
This analysis has then been further carried on by \citet{tripicco95,worthey96,trager98}, and
\citet{serven05}, urging in some cases a revision of the index label, as for the original Fe4668 
index \citep{worthey96,trager98}, 
now redefined as C$_2$4668 according to the real prevailing contribution in the evolutionary 
context of old stellar populations, as in elliptical galaxies.

The \citet{trager98} revision process has been intentionally left at the highest conservative
level, likely to avoid any massive re-definition of the original index set. However, we want 
to retake here some arguments of \citet*{worthey96} preliminary analysis to show that, at least 
for some relevant cases dealing with {\it bona fide} Fe features, 
an important update effort seems yet to impose. 
We will defer this delicate task to a forthcoming paper discussing here, just as an 
illustrative excercise, the case of index Fe4531.

\subsubsection{Absolute and net index sensitivity}

The main rationale for Lick-index set up \citep{worthey94b} aims at catching 
all the prominent absorption features in the optical spectra of late-type (type G-K) stars.
This uniquely requires the feature to be easily recognized at a $\sim 8.5$~\AA\ (FWHM) 
resolving power. If a blend could be envisaged for a feature, the latter is tacitly assumed
to be named after the element contributing with the deepest line, as recognized from the
list of theoretical atomic transitions \citep[e.g.][]{smith96,piskunov95,kupka99}. The deepest line 
also defines the feature centroid, which is reported in the \citet{worthey94b} original Table~1.
No special requirement is demanded to the two index side-bands, set to sample 
a nearby pseudo-continuum, other than being suitably ``feature-free'' and sufficiently extended in 
wavelength such as to be insensitive to stellar velocity dispersion broadening \citep{worthey94b}.

So, while the main focus is on the feature prominence, that is on feature's {\it absolute}
sensitivity to the ``leading'' chemical element, one should not neglect the fact that it is
the index ``net'' sensitivity (that is by netting the feature sensitivity to a given
element with the impact of the latter on the pseudo-continuum bands, too) that eventually 
set the real index response to a given chemical species.

\begin{figure*}
\includegraphics[width=0.47\hsize,clip=]{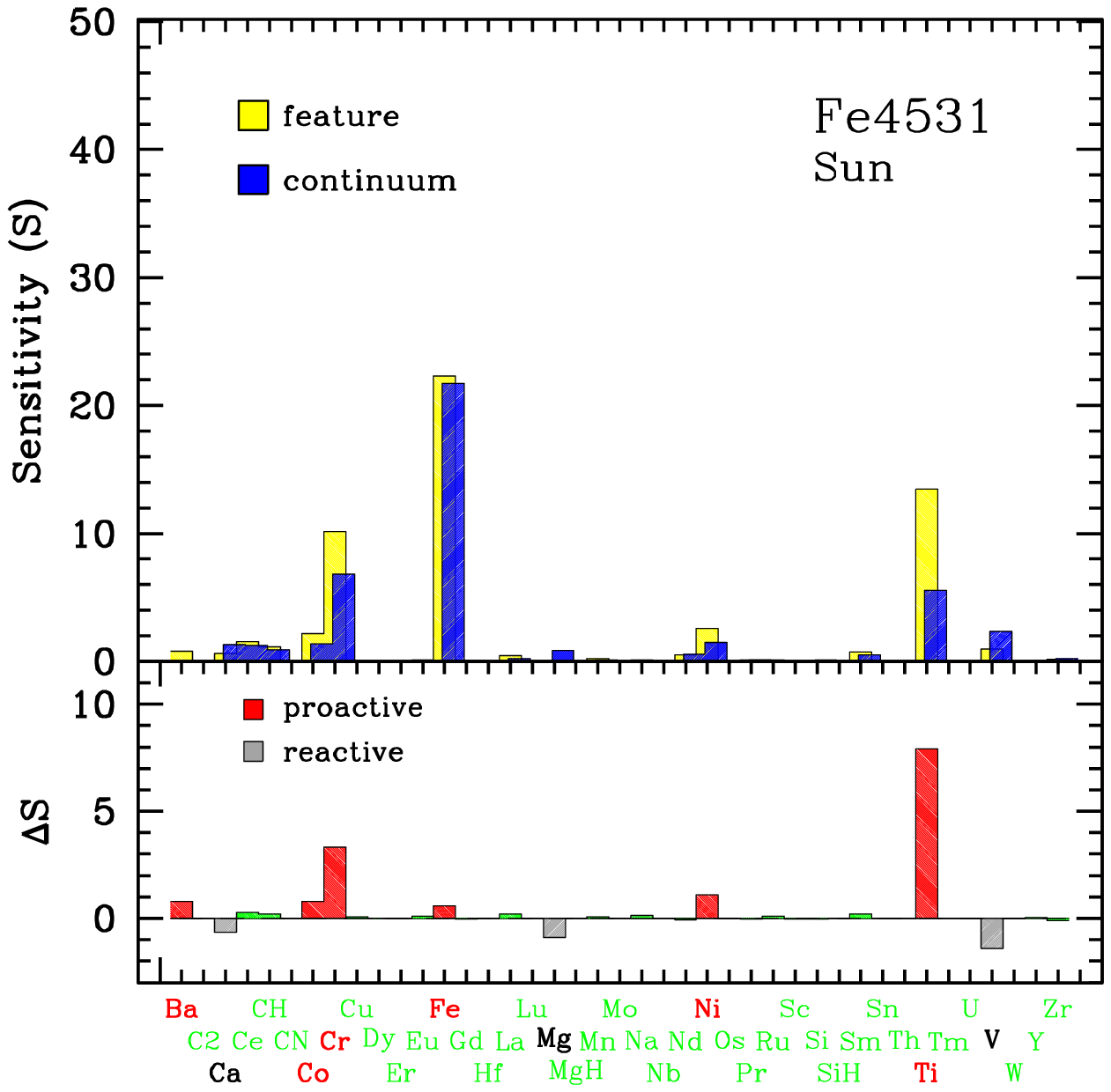}
\includegraphics[width=0.47\hsize,clip=]{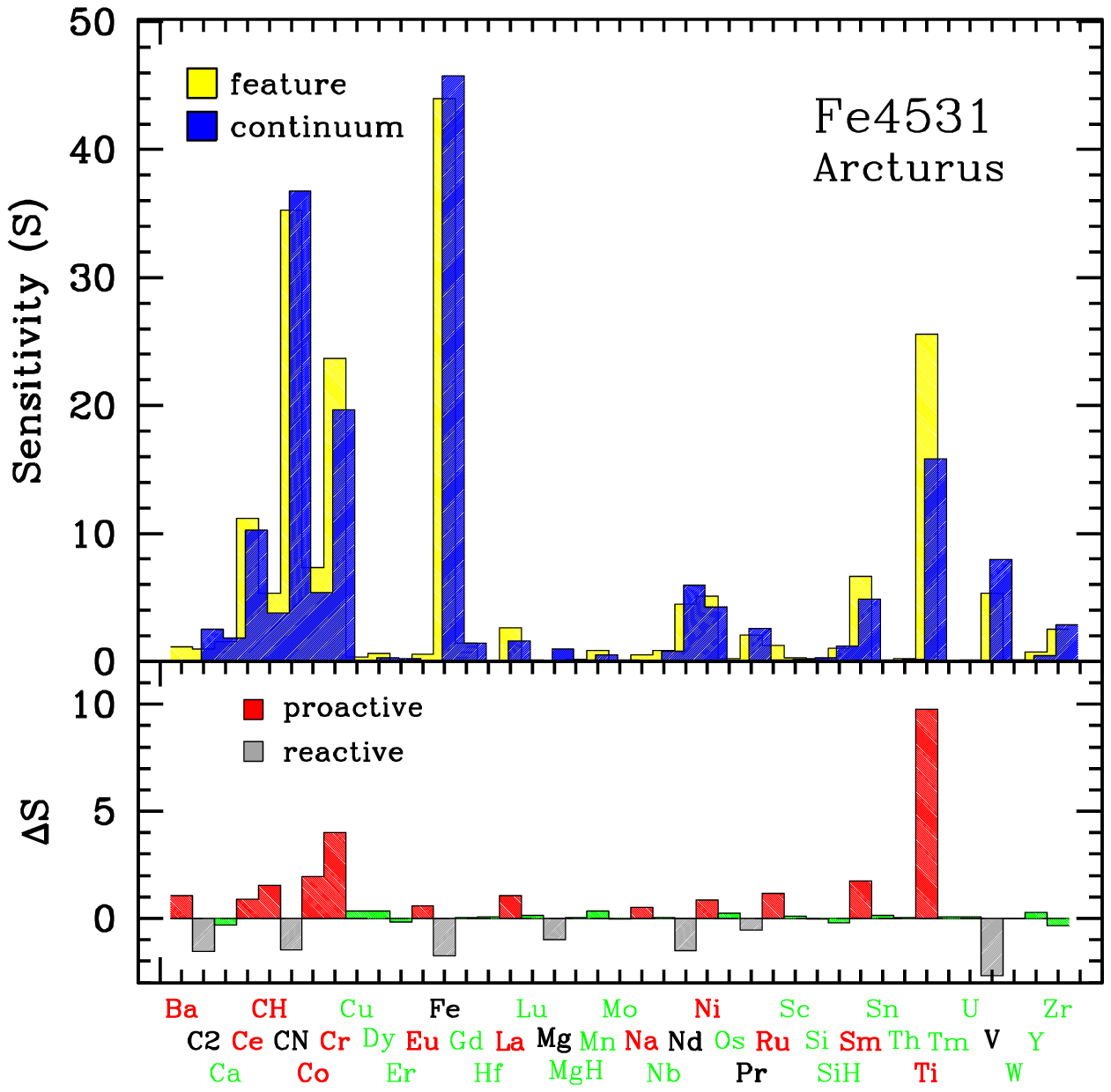}
\caption{
The Fe4531 index sensitivity to the 40 most prominent chemical elements (as sequenced in the x axis, 
disregarding their ionization states) in the Sun (left panel) and Arcturus (right panel) spectrum. 
In both panels, the yellow histograms refer to the elemental contribution (${\cal S}^f_X$) to the 
{\it feature} window, according to the definition of eq.~(\ref{eq:53}), while the blue histograms
are for the elemental contribution (${\cal S}^c_X$) to the flanking pseudo-continuum index sidebands,
according to eq.~(\ref{eq:54}). The {\it net} contribution to the index strength ($\Delta {\cal S}_X$),
as from eq.~(\ref{eq:55}), is displayed in the lower plots, where ``proactive'' elements are 
displayed in red and ``reactive'' elements in grey.
Note the leading role of Titanium and the reactive action of Vanadium and Magnesium.
See text for a full discussion.
}
\label{f15}
\end{figure*}

Like \citet{serven05}, in order to explore this important point we performed a test by probing the spectral
wavelength region around the Fe4531 feature in the high-res spectra of two reference
templates for old stellar populations, namely the Sun (type G2V, seen at $R\sim 350,000$
resolving power) and the star Arcturus (type K1.5III, observed at $R\sim 80,000$). 
For our analysis we relied on the {\sc Spectroweb 1.0} interactive database 
of spectral standard star atlases \citep{lobel07,lobel08,lobel11}. The Web program interface
allows an accurate fit of the observed spectra of these stars, supplying a detailed identification
of all the chemical species with absorption features within the wavelength range
(feature plus pseudo-continuum windows) of the Lick Fe4531 index. 

The sensitivity of both the 
feature- and the continuum windows to the first 40 most important absorbers has been quantified 
by summing up all the relevant line depths ($d = 1-I(\lambda)/I_c(\lambda)$) as from the 
high-res spectrum. As we always are dealing with weak (i.e. non-saturated) lines, one could assume, 
with sufficient accuracy, the equivalent width to simply scale with the line depth.
So, a measure of the absorbing strength of element $X$ (disregarding its 
different ionization states, in case) in the feature window of the index (of width $w$, 
as from Table~\ref{t1}) can be written as
\begin{equation}
{\cal S}^f_X = {\sum^w d_x}
\label{eq:53}
\end{equation}
The same can be done for the pseudo-continuum sensitivity to the corresponding element:
\begin{equation}
{\cal S}^c_X = \left({w \over W}\right) {\sum^W d_x},
\label{eq:54}
\end{equation}
where the width $W$ comes again from Table~\ref{t1}. As, in general, $w \neq W$, in 
the latter case ${\cal S}^c_X$ should be normalized by a factor $(w/W)$ to 
consistently compare with ${\cal S}^f_X$, as displayed in eq.~(\ref{eq:54}).

Clearly, the {\it net} sensitivity of the index to element $X$ is probed by the difference
\begin{equation}
\Delta {\cal S}_X = {\cal S}^f_X - {\cal S}^c_X,
\label{eq:55}
\end{equation}
or
\begin{equation}
{\Delta {\cal S}_X\over w} = {\sum^w d_x \over w}- {\sum^W d_x\over W}.
\end{equation}

If $\Delta {\cal S}_X >0$, then  element $X$ is ``proactive'' to the index strength (that is the 
index strengthens if $X$ abundance increases) otherwise, for $\Delta {\cal S}_X <0$,
$X$ is ``reactive'' (that is the index weakens if $X$ abundance increases, as $X$
more strongly absorbs the pseudo-continuum than the feature window). 
The results of our test on the Fe4531 index, are displayed in the two panels of Fig.~\ref{f15}, 
for the Sun and Arcturus, respectively. 

Accordingly, in Table~\ref{t6} we list the five most prominent absorbers to the feature 
window {\it alone} in the Sun and Arcturus.
These data definitely show that Fe is the strongest contributor to the 4531~\AA\ blend itself.
On the other hand, Fig.~\ref{f15} also show that Fe lines affect with a similar
strength also the pseudo-continuum flux density. When netting both contributions,
as in Table~\ref{t7}, one is left with a quite different list of prevailing chemical species that 
{\it really} settle the Fe5431 behaviour. In particular, {\it the main driver of the index
appears to be Titanium (!)}, followed to a lesser extent by Chromium. Note, in addition, the 
weak but steady inverse correlation of the index with the Vanadium abundance, which acts
therefore as a ``reactive'' element.

We could further carry on our analysis by probing the Fe4531 index response to homogeneous 
groups of chemical elements, like for instance the $\alpha$-element chain, more directly 
related to SNII nucleosynthesis \citep{woosley95}, (including the combined contribution of Ca, Mg, 
MgH, Na, Si, SiH), or the Fe-Ni group (including Co, Cr, Fe, Mn, Ni, V), as a main by-product 
of SNIa activity \citep{nomoto97}. This is summarized in the bottom lines of Table~\ref{t7}.
While Ti is definitely the main element that modulates Fe4531 strength, the other SNII $\alpha$
elements act as very weak ``reactive'' contributors, opposite to SNIa elemental group, which
``proactively'' contributes to the index strength, mainly in force of Cr, Co and Ni contributions.
Definitely, in spite of its outstanding r\^ole, our conclusion is that Fe plays
as a quite dull actor in this game.

\begin{table}
\caption{The five chemical species most important contributors to the
4531~\AA\ blend in the Sun and Arcturus spectra
}
\label{t6}
\begin{tabular}{lrrlr}
\hline\hline
\multicolumn{2}{c}{\hrulefill~Sun~\hrulefill} & & \multicolumn{2}{c}{\hrulefill~Arcturus~\hrulefill} \\ 
Chemical & &  & Chemical &  \\
species &  \multicolumn{1}{c}{${\cal S}^\dagger$} & & species & \multicolumn{1}{c}{${\cal S}^\dagger$} \\
\hline
Fe   & 22.3 &  &   Fe   &   44.0   \\ 
Ti   & 13.5 &  &   CN   &   35.3   \\ 
Cr   & 10.1 &  &   Ti   &   25.6   \\ 
Ni   &  2.6 &  &   Cr   &   23.7   \\ 
Co   &  2.2 &  &   Ce   &   11.2   \\ 
\hline\hline
\noalign{$^\dagger$ According to eq.~(\ref{eq:53}).}
\end{tabular}
\end{table}

\section{Discussion \& conclusions}

Since its definitive settlement, in the early 90ies \citep{gorgas93,worthey94b}, the Lick-index 
system has broadly imposed as the standard in the study of stars and old stellar systems. 
As a winning strategy, it linked the low-resolution spectroscopic approach to the narrow-band 
photometry such as to deliver at a time a condensed yet essential piece of information
about all the relevant absorption features that mark the SED of an 
astronomical target in the 4000-6000~\AA\ wavelength range. 

Nowadays, the Lick-index analysis consistently flanks other resources of modern
extragalactic astronomy at optical wavelength, including high-resolution spectral synthesis 
\citep[e.g.][]{vazdekis01,buzzoni05,rodriguez05,bertone08,vazdekis10,maraston11}
and low-resolution global fitting of galaxy spectral energy distribution (SED)
\citep[e.g.][]{massarotti01,bolzonella10,dominguez11,pforr12,gonzalez12,scoville13}.
However, as we have being discussing before, none of these is free from some limitations so that
narrow-band spectrophotometry still remains, in most cases, the only viable diagnostic 
tool to tackle with some confidence the evolutionary status of unresolved distant galaxies.

Along the original bulk of 21 indices, later on \citep{worthey97}, the Lick system has been complemented 
by including some higher-order Balmer lines 
(namely, H$\gamma$ and H$\delta$), with the aim of overcoming the possible perturbation of gas emission on 
the other optical features (especially H$\beta$) in the galaxy spectra. An alternative approach to 
the same problem was already attempted by \citet{gonzalez93}, to size up the H$\beta$ 
emission component in the spectra of active elliptical galaxies, by using [OIII] 5007~\AA\ emission 
as a proxy. 
Although nominally not part of the original Lick system (being the only index that explicitely covers
an emission feature), the \citet{gonzalez93} [OIII] index is often included in the extragalactic 
studies accompanying more classical Lick analysis
\citep[e.g.][]{kuntschner01,prugniel01,beuing02,nelan05,ogando08,wegner08}. To all extent,
we can consider it as a further addition to the standard system, which amounts, allover, to a total
of 26 indices (see their wavelength distribution in Fig.~\ref{f1}). 

A recognized difficulty, when using Lick indices from different observing environments, deals with the 
standardization procedure. As we have being discussing in Sec.~2.1, such a trouble comes out, from 
one hand, from the inherent difficulty with the Lick IDS set of primary stellar calibrators,
originally generated by the analogical output of a video camera tube \citep{robinson72}. As a result, IDS 
spectra lack any stable wavelength scale and intensity response, a limitation that, by itself, 
constrains calibration accuracy and prevents any consistent assessment of the spectra S/N ratio 
\citep{faber76,worthey94b}.

\begin{table}
\caption{The five chemical species with the most prominent {\it net} sensitivity
in the the Fe4531 index for the Sun and Arcturus spectra
}
\label{t7}
\begin{tabular}{lrrlr}
\hline\hline
\multicolumn{2}{c}{\hrulefill~Sun~\hrulefill} & & \multicolumn{2}{c}{\hrulefill~Arcturus~\hrulefill} \\ 
Chemical & &  & Chemical  & \\
species & \multicolumn{1}{c}{$\Delta {\cal S}^\dagger$}   & & species & \multicolumn{1}{c}{$\Delta {\cal S}^\dagger$}  \\
\hline
Ti   &    7.9  & &   Ti   &    9.8   \\  
Cr   &    3.3  & &   Cr   &    4.0   \\  
V    &  --1.4  & &   V    &  --2.7   \\  
Ni   &    1.1  & &   Co   &    1.9   \\  
Mg   &  --0.9  & &   Sm   &    1.8   \\  
\hline
SNIa group &    4.4  & &        &    2.7   \\
SNII group &  --1.4  & &        &  --1.0   \\
\hline\hline
\noalign{$^\dagger$ According to eq.~(\ref{eq:55}).}
\end{tabular}
\end{table}

On the other hand, the problem of index standardization intimately relates to a tricky and yet not fully 
realized theoretical drawback. In fact, even if both feature ($f$) and pseudo-continuum ($f_c$) density 
fluxes, as of eq.~(\ref{eq:r}),
obey a normal distribution, the ${\cal R} = \overline{(f/f_c)}$ ratio may not.\footnote{In some 
cases, the $\cal R$ ratio may even display a bi-modal statistical distribution, 
\citep[see][]{marsiglia64,marsiglia65}.}
The index definition, in terms of ratio ${\cal R}$, is a direct legacy of the ``spectroscopic mood'' 
of the Lick system, and it closely recalls the standard procedure to derive line equivalent width as in 
high-resolution spectra. Given the different conditions of low-resolution unfluxed spectra, however, we 
have demonstrated that a somewhat alternate ``photometric mood'' should be preferred, with feature strength 
assessed in terms of ratio ${\cal R}' = (\overline{f}/\overline{f_c})$. The latter is a far more suitable 
and robust statistical estimator, which displays a Gaussian distribution and turns index strength in terms 
of a narrow-band ``color''. Under certain conditions, we have shown that ${\cal R}$ tends to ${\cal R}'$ 
{\it providing a spectrum is nearly ``flat'' and overcomes a minimum S/N threshold, such as}
\begin{equation}
(S/N)_{obs} \ga\  22 \left({\theta \over {\cal W}}\right)^{1/2} \, \approx \,5\,\theta^{1/2} \quad [px^{-1}],
\label{eq:snmin}
\end{equation}
being $\theta$ the spectral dispersion (that is the {\it observed}  wavelength pixel scale), 
and ${\cal W}$ the average index window size as from eq.~(\ref{eq:d}) and Table~\ref{t1}.
This means, for example, that for a resolving power $R \sim 2000$, a minimum 
$(S/N)_{obs} \sim 5$~px$^{-1}$ ratio has to be reached, assuming a Bayesian pixel fair sampling.

{\it A relevant conclusion of our analysis is therefore that, in an effort to revise the Lick 
system, one should better consider to compute indices either in the classical form of
pseudo-equivalent width, relying on the classical ${\cal R}$ ratio of eq.~(\ref{eq:r}), but
from previously linearized spectra according to eq.~(\ref{eq:fmax}), or (more safely) in the form 
of ``narrow-band'' colors of simply fluxed spectra, that is in terms of the ${\cal R}'$ ratio of 
eq.~(\ref{eq:r1}), and express them in magnitude scale, as from eq.~(\ref{eq:imag}).}

As a further issue in our analysis, in Sec.~3 we assessed the possible redundancy in the index
delivered information when matching for instance galaxy spectra. About 25\% of the covered wavelength
range, in fact, contributes to the definition of two or more Lick indices, a feature that 
may lead index changes to correlate, in some cases, beyond any strictly physical relationship. 
A relevant case has been explored, for instance, dealing 
with the nested configuration of the Fe5335, Fe5270 and Mg$_2$ indices, a classical triad often involved 
in literature studies to define ``global'' metallicity indices, such as the 
$\langle Fe \rangle = (Fe5270+Fe5335)/2$ of \citet{gorgas90}, or the perused 
$[MgFe] = [{{\rm Mgb}\,(Fe5270+Fe5335)/2}]^{1/2}$ meta-index of \citet{gonzalez93}, in its manifold variants 
\citep[see, e.g.][]{fritze98,thomas03,clemens06}.
We have shown, in Sec.~3.4, that any intervening change in the Fe5335 strength also reflects in 
the Mg$_2$ index (and in Mg$_1$, as well) in quite a puzzling way. From one hand, if a
flux addition increases the pseudo-continuum level in common, then both Mg$_2$ and Fe5335 strengths
are seen to increase, according to eq.~(\ref{eq:48}). On the other hand, if the Fe5335 index becomes
stronger due to a deeper corresponding feature, then Mg$_2$ is seen to decrease, as from 
eq.~(\ref{eq:50}).

To further embroil the situation, in Sec.~3.3 we made a point dealing with the the Mgb index as a tracer of 
the galaxy metallicity, for instance within the [MgFe] meta-index.
As shown by \citet{goudfrooij96}, the strength of the Mg atomic feature itself could be tacitely
modulated by the intervening gas emission, even in relatively quiescent stellar environments. The forbidden 
[NI] line emission about 5199~\AA\ adds flux to the red Mgb pseudo-continuum (see the sketch in 
Fig.~\ref{f10}), thus leading to a stronger Mg absorption feature, overall.
In Fig.~\ref{f12b} the effect has been clearly detected on a statistical basis by relying on the \citet{ogando08} 
extended galaxy database. The [NI] emission is seen to correlate with the [OIII] strength, roughly in a
ratio 1:5, and it appears to be in excess of about 0.4~\AA\ among the most ``active'' (i.e.\ [OIII]-strong) 
objects in the sample compared with the quiescent ellipticals. According to eq.~(\ref{eq:n3}), the Mgb strength 
is increased by a similar figure. When relying, for instance, on the \citet{pipino11} empirical calibration 
to convert Lick indices into metal abundance (see, in particular Table~5, therein), {\it this leads to 
predict a slightly higher metallicity, even in excess of some 0.2~dex in star-forming galaxy environments.} 

\begin{figure*}
\includegraphics[width=0.9\hsize,clip=]{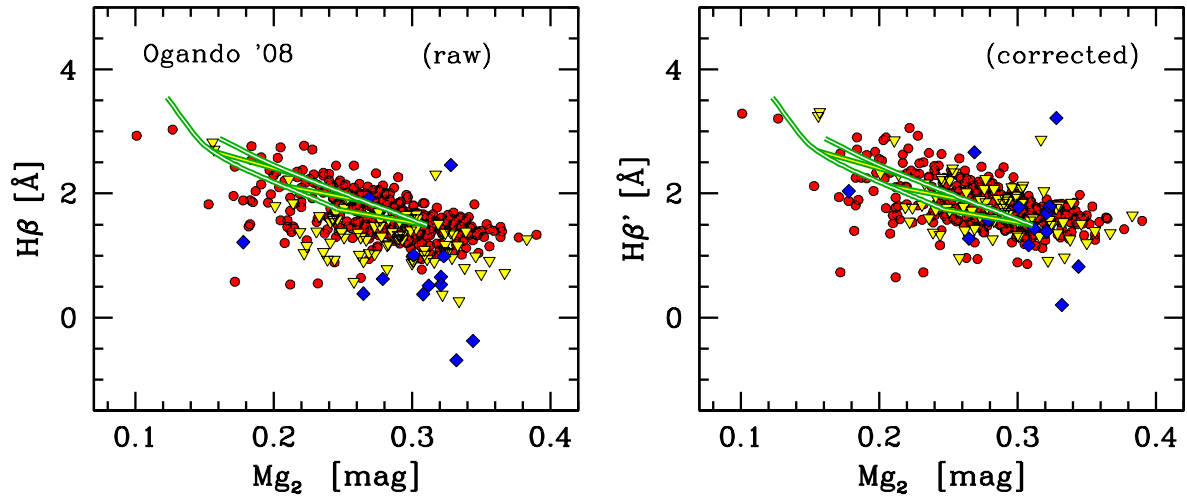}
\caption{
The devised scheme for H$\beta$ correction, as in Fig.~\ref{f9}, is applied to the \citet{ogando08} early-type
galaxy sample, displayed in the H$\beta$ versus Mg$_2$ index plane. The strength of H$\beta_e$ 
gas emission is sized up and subtracted according to eq.~(\ref{eq:33}), taking the observed
[OIII]'$_{5007}$ emission as a proxy. To better appreciate the correction effect, we adopted three
different galaxy markers for nearly ``quiescent''  (i.e.\ [OIII]'$_{5007} > -1$~\AA, or 
H$\beta_e \ga -0.4$~\AA, solid dots), ``moderately active'' ($-2 \le $[OIII]'$_{5007} \le -1$~\AA, 
or $-1.5 \la $ H$\beta_e  \la -0.4$~\AA, triangles), and ``active'' (i.e.\ [OIII]'$_{5007} < -2$~\AA\ 
or H$\beta_e \la -1.5$~\AA, rombs) galaxies. As a guideline, in both panels we also displayed
the \citet{buzzoni94} synthesis models for a standard Salpeter IMF and [Fe/H] metallicity parameter
between $-0.25$ (upper line) and $+0.22$~dex (lower line). ``Ladders'' across model sequences tick
galaxy age for values of 2, 5, 10, and 15 Gyr, with increasing Mg$_2$. 
After correction (right panel), galaxy H$\beta$ index is increased, on average, by $0.25 \pm 0.20$~\AA\ 
along the whole sample, making galaxies to appear, on average, about 20-30\% younger and 0.2-0.3~dex
metal poorer. 
}
\label{f18}
\end{figure*}

Like the [NI] case, hidden emission can subtly plague the nominal diagnostic scheme for other Lick indices.
Indeed, the most debated issue, in this regard, deals with a fair correction of H$\beta$ absorption
in the spectra of elliptical galaxies. Compared to all the other Lick indices, 
H$\beta$ \citep[and its weaker Balmer companions, like H$\gamma$ and H$\delta$, see][]{worthey97} is the only 
feature that positively correlates with the effective temperature of stars \citep[see, e.g.\ Fig. 4 in][]{trager98}.
As it becomes strongest among A-type stars, this makes the index of special interest to break model degeneracy 
in the age diagnostic of old- and intermediate- ($t\la 1$~Gyr) age galaxies 
\citep{buzzoni94,worthey94,maraston00,tantalo04}.
On the other hand, as gas and stars within a galaxy are conflicting in their line contribution (the first, by
emitting, the latter ones by absorbing), the integrated H$\beta$ strength displays the net effect of this 
interplay. 

In Sec.~3.2 we tackled the \citet{gonzalez93} well experienced  correction procedure, that relies on the 
measure of the adjacent [OIII] forbidden emission at 5007~\AA, as a proxy to infer the H$\beta$ gaseous 
contribution. Several physical arguments can be advocated to question 
such a straightforward relationship between the two emission lines, in consequence of a range of 
different evolutionary mechanisms ({\it in primis} LINER activity) at work. 
However, one can admittedly assume [OIII] emission to be a generic (but not exclusive) tracer of 
star-forming activity (and the implied presence of HII gas clouds) within a young stellar population.

Taking a set of spectroscopic observations of 20 bright elliptical galaxies, originally aimed at studying
radial spectrophotometric gradients \citep{carrasco95,buzzoni09b,buzzoni15}, we devised a simple and robust diagnostic
scheme to probe the possible [OIII] versus H$\beta$ entanglement. As observations were taken symmetrically
across the galaxy major axis, one can reasonably assume that, whatever the distinctive evolutionary properties,
the same stellar population (and, on average, gas content) is sampled at the same radial distance, on opposite 
sides from the galaxy center. If this is the case, then the H$\beta$ and [OIII]$_{5007}$ index differences 
between the two galaxy sides can be contrasted in search for possible statistically correlations. One major 
advantage of this approach is that it is nominally free from zero-point and other calibration problems with 
the data, just relying on the assumed central symmetry of the galaxy stellar population.

{\it Our results, summarized in Fig.~\ref{f8} and Table~\ref{t5}, basically confirm \citet*{gonzalez93} correction
figure of $\Delta H\beta/\Delta[OIII]_{5007} \sim 0.7 \pm 0.5$, but only for those 
ellipticals with established on-going star-formation activity and stronger core emission 
(i.e.\ $[OIII]_{5007} \la -1.5$~\AA).} As far as more quiescent galaxies are considered, any H$\beta$ 
emission is found to fade out more quickly than [OIII]$_{5007}$ emission, and this suggests a milder correction 
scheme, that we parameterized by means of a logistic curve (see Fig.~\ref{f9}), as in eq.~(\ref{eq:33a}) and (\ref{eq:33}).

The effect of our H$\beta$ correction can be appreciated by looking again at the \citet{ogando08} 
sample of elliptical galaxies, as in Fig.~\ref{f18}. The corrected galaxy distribution in the H$\beta$ versus 
Mg$_2$ plane is shown in the right panel of the figure, compared with the original sample, as in the left
panel. Just as a guideline, the \citet{buzzoni94} models are overplotted, throughout. 
One can notice that most of those galaxies with ``shallow'' (or negative) H$\beta$ in the left panel are 
effectively recovered by the procedure, leading eventually to a much cleaner H$\beta$ versus Mg$_2$ relationship. 
Along the whole galaxy sample, the corrected H$\beta$ index slightly increases, on average, by  
$\Delta H\beta \simeq 0.25 \pm 0.20$~\AA. Although apparently negligible, {\it this figure has 
important impacts on the inferred galaxy age, which typically shifts toward 20-30\% younger values 
with a 0.2-0.3~dex poorer inferred metallicity} \citep{buzzoni94,worthey94}.

Aside the H$\beta$ problem, the striking importance of the [OIII]$_{5007}$ forbidden line also emerged
along our discussion on Sec. 3.2 when dealing with the intrinsic strength of the Fe5015 index. From one
hand, any intervening [OIII]$_{5007}$ emission will decrease Fe5015 strength as it adds flux to the 
feature window of the latter. On the other hand, the broader and stronger Fe5015 absorption 
in the integrated spectrum of old stellar populations may artificially depress the apparent 
[OIII]$_{5007}$ luminosity, thus leading to underestimate its equivalent width. In our analysis, we tackled 
in some detail the problem of a mutual correction of the two indices devising a general procedure that could 
straightforwardly be applied to other similar contexts, as well. The intrinsic [OIII]$_{5007}$ strength can 
therefore be derived in terms of observed quantities, via eq.(\ref{eq:30}), while the intrinsic Fe5015 index, 
corrected for O-emission, simply derives from eq.~(\ref{eq:27}). 

One important consequence of our analysis is that corrected Fe5015 index is generally expected to increase 
in the galaxy spectrum, opposite to Mgb trend. 
This feature may have some impact when combining the two indices, like in the [MgFe50] meta-index, so 
extensively used for instance in the {\sc Sauron} studies \citep{kuntschner10,falcon11}. 
An intersting comparison can be done, in this regard, with \citet[][see, in particular, Fig. 3, therein]{peletier07}.
In the plot, when taking the \citet{thomas03} stellar population synthesis models as a reference
in the Fe5015 vs.\ Mgb domain, it appears that {\sc Sauron} ellipticals better match
the $\alpha$-enhanced models with respect to the standard solar chemical partition.
On the other hand, one may argue that this is the natural consequence of a relatively ``strong'' Mgb 
and ``shallow'' Fe5015 so that, when accounting for the envisaged corrections, the net effect 
on the fit {\it is to mitigate any inferred $\alpha$-element enhancement of the galaxy population.}

For its relevance, the Mg index complex (including Mgb, Mg$_1$, and Mg$_2$ indices) has been the focus
of our specific analysis in Sec.~3.3. In its original intention, the Lick system aimed at singling out
the atomic and molecular contribution to the broad ``Mg valley'' about 5150~\AA\ \citep{faber77}.
This evidently leads to conceive some entanglement among the three indices, that we theoretically assessed, 
for the first time, by means of eq.~(\ref{eq:mg}). This inherent relationship can be usefully relied upon 
in case of incomplete spectral observations, either to derive the missing index (within a  $\sim$1.5\% internal 
accuracy) given the other two or, even more importantly, to assess any possible deviation from the standard scheme, 
as due to intervening bias effects in the spectra. This has actually been our approach, for instance, to single 
out the [NI]$_{5199}$ contribution, as shown in Fig.~\ref{f12b}.

As far as low-resolution spectra are dealt with, as in any Lick 
analysis, we are left with the unescapable consequence that any absorption feature is in fact a blend of a 
coarse elemental contribution.
As we have shown in Sec. 3.5, this problem has been widely explored, from a theoretical point of view, in 
the recent literature \citep[the work of][is probably the most up-to-date attempt in this sense]{serven05}
in order to assess index ``responsiveness'' to the different chemical species that may potentially intervene
at the relevant wavelength of the blended absorption feature. On the other hand, one has to notice in this
regard that, for its inherent definition, Lick-index strength directly stems from the {\it residual} 
``in-band'' elemental contribution to the feature itself, compensated by the corresponding ``off-band''
contribution to the flanking pseudo-continuum. Contrary to high-resolution spectroscopy, therefore, a 
tricky situation may actually set in, where 
the Lick-index christening element (that is the one with the deepest absorption within the feature window) 
is not necessarily the representative one that constrains the index strength, as a whole.

As a recognized example \citep{trager98,serven05}, in Sec. 3.5 we retook in somewhat finer detail the
case of Fe4531, one among the many {\it bona fide} Iron tracers in the optical spectrum 
of galaxies, according to the classical Lick scheme. For our analysis we relied on the Sun and Arcturus 
observed spectra as typical main-sequence and red-giant stellar templates, in an attempt to extend
our conclusions also to the old-galaxy framework, according to more elaborated stellar 
population synthesis models.
The comparison of Table~\ref{t6} and \ref{t7} is illuminating to catch the sense of our analysis.

One sees, in fact (Table~\ref{t6}) that, both in the Sun and Arcturus (and, by inference, in the spectrum 
of elliptical galaxies, for instance), {\it Iron} is undoubtedly the leading contributor to the 
absorption feature at 4531~\AA, but once clearing the Fe coarse absorption in the nearby 
pseudo-continuum as well, it is actually {\it Titanium} that, almost solely, commands the game 
(Table~\ref{t7}) and eventually modulates Fe4531 index behaviour.
As a final consideration, the case of Fe4531 and more generally the essence of all our previous discussion, 
evidently urge an important rethinking process of the Lick-index tool, to bring it fresh life and even
better tuned diagnostic performance in view of its updated application to the new-generation spectral data
to come.

\section*{Acknowledgments}

The anonymous referee is warmly acknowledged for his/her competent and very careful review of the draft,
and for a number of timely suggestions, all of them of special relevance to better focus our analysis.
This work has made extensive use of the {\sc Spectroweb 1.0} interactive database of spectral standard star 
atlases, maintained by Alex Lobel and collaborators at the Royal Observatory of Belgium, in Brussel.

\end{document}